\documentclass[aps,prl,reprint, superscriptaddress]{revtex4-2}
\usepackage{soul}
\usepackage{lineno}
\usepackage{graphicx}
\usepackage{dcolumn}   
\usepackage{bm}        
\usepackage{amssymb}   
\usepackage{dsfont}
\usepackage{relsize}
\usepackage{amsmath,stmaryrd}
\usepackage{blkarray, multirow, graphicx, diagbox, color, colortbl}
\usepackage{bbm} 
\usepackage{glossaries}
\usepackage{hyphenat}
\usepackage{ifthen}
\usepackage[colorlinks, linkcolor = blue, citecolor = blue, filecolor = black, urlcolor = blue]{hyperref}
\usepackage{xkeyval}
\usepackage{moreverb}
\usepackage{rotating}
\usepackage{wrapfig}
\usepackage{slashbox}
\usepackage{xspace}
\usepackage{nicefrac}
\usepackage[]{units}
\usepackage{physics}
\usepackage{booktabs}
\usepackage{braket}
\usepackage[inline]{enumitem}
\usepackage{tabto}
\usepackage{listings}
\usepackage{xstring}
\usepackage{url}
\usepackage[section]{placeins}
\usepackage[normalem]{ulem}
\usepackage{mathrsfs}
\usepackage{mathtools}
\usepackage[scr=boondoxo]{mathalpha}
\usepackage[percent]{overpic}

\def\ReplaceStr#1{%
	\IfSubStr{#1}{p}{%
		\StrSubstitute{#1}{p}{.}}{#1}}

\usepackage[capitalise]{cleveref} %
\usepackage{chemformula}
\usepackage{algorithm}

\usepackage{tikz}

\usetikzlibrary{calc}

\usepackage[dvipsnames]{xcolor}
\definecolor{mygreen}{HTML}{009000}
\definecolor{myorange}{HTML}{ff6600}
\definecolor{mypurple}{HTML}{6F00FF}
\definecolor{electriclime}{rgb}{0.8, 1.0, 0.0}

\definecolor{block}{RGB}{0,162,232}

\usepackage{tikz}

\definecolor{cibord}{HTML}{ff7f31}   
\definecolor{cifill}{HTML}{fff7f3}      

\usetikzlibrary{shapes.geometric}

\definecolor{bluebord}{RGB}{63,79,162}  
\definecolor{bluefill}{RGB}{214,215,236} 

\definecolor{sqbord}{HTML}{8d6bb8}
\definecolor{sqfill}{HTML}{f4f2f4}

\definecolor{tribord}{HTML}{529d3f}
\definecolor{trifill}{HTML}{f5f7f4}

\def\blockaux#1(#2,#3)#4(#5,#6){%
  \draw[fill={#1}, draw=none]
  let \p1=(#2,#3),
      \p2=(#5,#6),
      \p3=(#2+#5,#3+#6),
      \p4=(#2+#5/2,#3+#6/2)
  in
    (\p1) rectangle (\p3)
    (\p4) node {\centering $#4$}
  ;%
}

\newacronym{OBC}{OBC}{open boundary condition}
\newacronym{PBC}{PBC}{periodic boundary condition}
\graphicspath{{figures/}}
\def\affilA{Department of Physics, Arnold Sommerfeld Center for Theoretical Physics (ASC), Munich Center for Quantum Science and Technology (MCQST), Ludwig-Maximilians-Universit\"{a}t M\"{u}nchen, 80333 M\"{u}nchen, Germany}
\def\affilB{Université Paris-Saclay, CNRS, LPTMS, 91405, Orsay, France}
\def\affilC{School of Physics, Trinity College Dublin, Dublin 2, Ireland} 
\def\affilD{Rinn Quantum, School of Physics, Trinity College Dublin}

\begin{document}
\def\thetitle{Locality-Preparation Time Tradeoffs in Dissipative Quantum Systems} 
\author{Philipp Westhoff}\email{p.westhoff@physik.uni-muenchen.de}
\affiliation{\affilA}
\author{Sara Murciano}\email{sara.murciano@universite-paris-saclay.fr}
\affiliation{\affilB}
\author{Mattia Moroder}
\email{moroderm@tcd.ie}
\affiliation{\affilC}
\affiliation{\affilD}

\title{\thetitle}
\begin{abstract}
Dissipative state preparation engineers quantum many-body states as fixed points of an open-system dynamics. The spatial range of the required dissipation sets both the feasibility of analog implementations and the connectivity and control complexity of digital ones. Yet how this locality quantitatively constrains preparation times remains largely unknown. Here, we establish an exact locality-preparation time relation for quadratic fermionic systems and develop a construction that optimizes the locality of the dissipative dynamics while preserving the target state. Applied to the Kitaev chain, our framework reveals sharply different locality-preparation time tradeoffs in critical and gapped regimes. At criticality, suppressing the algebraic tails of the jump operators necessarily forces the preparation time to grow algebraically with system size. By contrast, both trivial and topological gapped phases admit exponentially localized dissipation with a finite preparation time in the thermodynamic limit. Our results identify when fast dissipative preparation is compatible with local control and when criticality imposes an unavoidable dynamical cost.
\end{abstract}
\maketitle
Engineering complex quantum states in many-body systems is a central challenge in quantum computation and simulation~\cite{Bloch2012-af, Georgescu2014, Altman2021, eisert2025mindgaps}. 
Existing approaches include adiabatic~\cite{Sokolov2024} and quench~\cite{Bernier2025} protocols, variational~\cite{Wang2022bang} and tensor-network-inspired methods~\cite{Haghshenas_2024}, and measurement- or feedback-assisted schemes~\cite{Lu_2022,Swain2025}.
Dissipative state preparation (DSP) provides a complementary strategy, in which engineered system-environment interactions make the target state an attractive fixed point of the dynamics~\cite{Diehl2008, Kraus2008, Verstraete2009, Diehl_2011, PRXQuantum.3.010301, Mi2024, Schnell2024, Chen2025-fv, Zhan2025, 4hx7-xnhw, Ong2026, Chowdhury2026, schloemer2026}. This fixed-point structure can reduce sensitivity to initialization and provide resilience against imperfections~\cite{Lin2025advantage}.
\begin{figure}[h!]
    {\centering
    \begin{overpic}[width=\columnwidth]{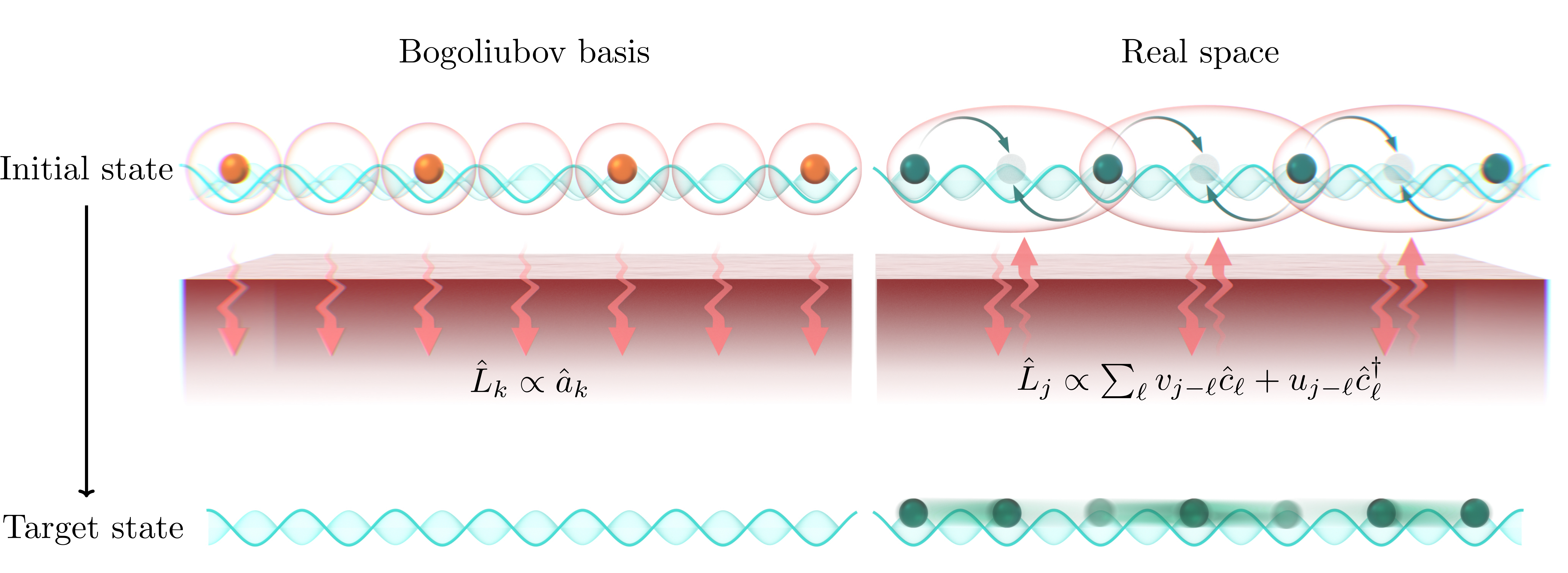}
        \put(0,30){(a)}
    \end{overpic}\\
    \vspace{0.1cm}
    \begin{overpic}[width=0.95\columnwidth]{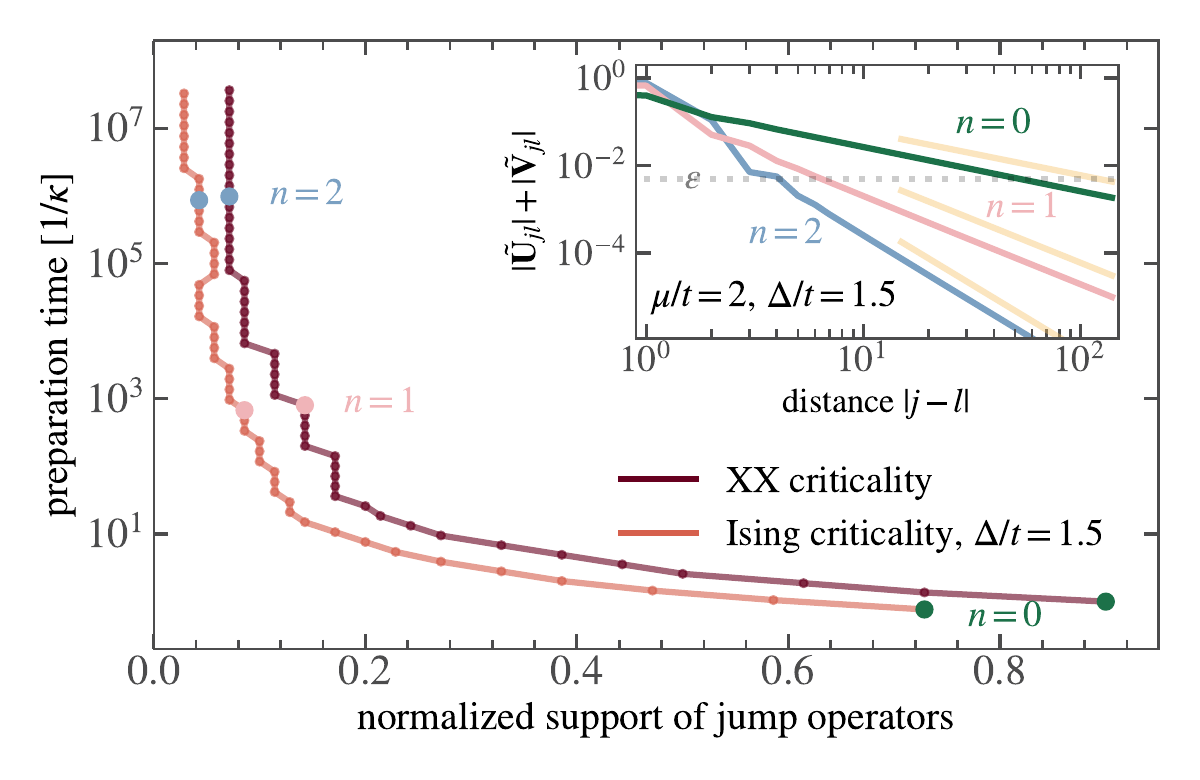}
        \put(-9,60){(b)}
    \end{overpic}\par}
    \caption{
\textbf{Locality-preparation time tradeoff.}
(a) Schematic of dissipative ground state preparation. In the Bogoliubov basis (left), the target ground state is simply the quasiparticle vacuum, whereas in real space (right) it has a nontrivial structure. In the main text, we show how the jump operators can be optimized to maximize their real-space locality while preserving the target steady state.
(b) Preparation time, expressed in units of $\kappa^{-1}$, as a function of the normalized support of the jump operators at XX and Ising criticality (see~\cref{fig:phase:diagram}). Highlighted points correspond to integer values of the suppression exponent $n$. The inset shows the corresponding real-space decay of the jump-operator coefficients at Ising criticality, with the yellow lines indicating the expected algebraic decay $\sim |j-l|^{-(n+1)}$. Data are shown for a system of $N=140$ lattice sites, using a support cutoff $\epsilon=0.005$.
}
    \label{fig:first}
\end{figure}

A central question in DSP concerns the fundamental limits on preparation times. Local Lindbladians obey Lieb-Robinson bounds~\cite{Poulin2010}, which constrain the propagation of information and therefore the build-up of long-range correlations.
Moreover, it has been shown that local gapped Lindbladians feature exponential decay of correlations in the steady state~\cite{Kastoryano_2013}.
Complementarily, recent work has established a fundamental tradeoff between steady-state entanglement and relaxation time under bipartite locality constraints~\cite{Pocklington2024Time-Ent-Tradeoff}.
Together, these results show that locality limits DSP, but leave open how the preparation time depends on the spatial range of the jump operators. Quantifying this dependence is essential for scalable implementations, where the jump operator range sets the required connectivity and control complexity.

One-dimensional topological superconductors provide a minimal setting in which this question can be addressed across different phases. Their topological phase hosts Majorana boundary modes and supports nonlocal information encoding~\cite{Kitaev2001,Nayak2008}. These systems also host critical states with algebraically decaying
correlations, central to many-body physics~\cite{sachdev_quantum_2011,sachdev_gapless_1993}
and quantum simulation~\cite{Simon2011-zh,keesling_quantum_2019,ebadi_quantum_2021,manovitz_quantum_2025,andersen_thermalization_2025,fang_probing_2024,Sun2026experimental,Mo2026observing},
with potential applications in decoherence-resilient
metrology~\cite{Chen2026,Musso2026}.
Their divergent correlation length, however, makes them especially
challenging to prepare~\cite{Desales2026}.
This raises a central question: do criticality and topology impose
a fundamental tradeoff between dissipator locality and preparation time?

Here, we develop a general framework for quantifying
locality-preparation time tradeoffs in quadratic fermionic systems.
The preparation strategy is illustrated in~\cref{fig:first}(a): the system is driven toward the ground state by dissipating quasiparticles.
We exploit the freedom to mix Bogoliubov annihilation operators without changing the target state, constructing jump operators that are optimally localized in real space and determining the associated preparation-time cost. 
Applying this framework to the Kitaev chain, we establish our main
result: an exact locality-preparation time tradeoff at criticality.
Faster algebraic decay of the jump operators necessarily entails
a larger exponent governing the growth of preparation time with
system size.
By contrast, in both trivial and topological gapped phases, exponentially localized jump operators can coexist with a finite preparation time in the thermodynamic limit. Moreover, along a special curve in the topological phase, the optimal jump operators become strictly finite-range. 

\paragraph{Optimally localized dissipators--}
Any quadratic fermionic Hamiltonian can be written in the form
\begin{equation}
    \hat{H} = \sum_k \varepsilon_k\, \hat{a}_k^\dagger \hat{a}_k,
\end{equation}
where $\hat{a}_k^\dagger$ ($\hat{a}_k$) creates (annihilates) a Bogoliubov quasiparticle and by an appropriate particle-hole convention, we can restrict to
$\varepsilon_k\geq0$ without loss of generality~\cite{Kitaev2001}. 
We consider Markovian dissipative processes obeying~\cite{Lindblad1976b}
\begin{equation}
 \frac{d}{dt} \hat\rho = -i [\hat H, \hat\rho] + \sum_j \mathcal{D}_{\hat L_j}[\hat\rho] \equiv \mathcal{L}[\hat \rho]\,, 
 \label{eq:lindblad}
\end{equation}
where $\hat\rho$ is the density matrix of the system, $\mathcal{D}_{\hat L_j}[\hat\rho] = \hat L_j \hat\rho \hat L_j^\dagger - \frac{1}{2}\{\hat L_j^\dagger \hat L_j, \hat\rho\}$ is the dissipator associated with jump operator $\hat L_j$ and $\mathcal{L}$ is the Lindbladian generating the dynamics.
With the above convention, the target ground state is the Bogoliubov vacuum $|\Omega\rangle$, satisfying $\hat a_k |\Omega\rangle = 0$ for all $k$. Targeting this state dissipatively is then straightforward: the jump operators $\hat L_k = \sqrt{\kappa}\,\hat a_k$, with dissipation strength $\kappa$, drive any initial state toward $|\Omega\rangle$, since each $\hat L_k$ continuously removes excitations from mode $k$.
If zero modes are present, the ground state is degenerate and we fix a target state within the ground-state manifold (see Supplemental Material (SM)). In the following we consider purely dissipative dynamics.

A practical limitation of this approach is that the Bogoliubov operators 
$\hat{a}_k$ are generally highly non-local in real space, making them 
difficult to realize experimentally. This limitation can be overcome by noting that the steady state is preserved under a broader class of jump operators
\begin{equation}
\hat L_j=\sqrt{\kappa}\sum_k \mathbf{B}_{jk}\hat a_k \,,  
\label{eq:general:jump:ops}
\end{equation}
where $\mathbf B$ is a full-rank mixing matrix, as we prove in the SM.
Since an overall rescaling of $\mathbf B$ changes only the dissipative timescale, we fix
$\|\mathbf B\|_{\mathrm{op}}=1$, so that $\kappa$ sets the overall rate.
Different choices of the mixing matrix can then be used to optimize the real-space locality of the jump operators.

\paragraph{Locality-preparation time tradeoff}
To link the matrix elements of $\mathbf B$ with the locality of the jump
operators, we express the Bogoliubov quasiparticles in terms of the physical
real-space fermionic operators 
\begin{equation}
    \hat a_k=\sum_{\ell}
\left(\mathbf{V}_{k\ell}\hat c_{\ell}
+\mathbf{U}_{k\ell}\hat c_{\ell}^{\dagger}\right) \,,
    \label{eq:bogoliubov:annihilator:from:real:space:operators}
\end{equation}
where $\mathbf V$ and $\mathbf U$ are the Bogoliubov transformation matrices. Substituting into~\cref{eq:general:jump:ops}, the mixed jump operators take
the form
\begin{equation}
   \hat L_j=\sqrt{\kappa}\sum_{\ell}
\left(\tilde{\mathbf{V}}_{j\ell}\hat c_{\ell}
+\tilde{\mathbf{U}}_{j\ell}\hat c_{\ell}^{\dagger}\right) \, ,
\label{eq:jump:operators:real}
\end{equation}
so that the locality of the jump operators depends on the asymptotic properties of the coefficients $\tilde{\mathbf{V}}_{j\ell}=\sum_k \mathbf{B}_{jk}\mathbf{V}_{k\ell}$ and $\tilde{\mathbf{U}}_{j\ell}=\sum_k \mathbf{B}_{jk} \mathbf{U}_{k\ell}$.
The relaxation timescale of the system solely depends on the mixing matrix $\mathbf{B}$, and can be extracted through third quantization~\cite{Prosen_2008}.

In the case of \glspl{PBC}, translation invariance of the jump operators constrains the mixing matrices to
\begin{equation}
\mathbf{B}_{jk}=\frac{e^{-ikj}}{\sqrt N}\,\beta_k \,, 
\label{eq:mixing:matrix}
\end{equation}
where $\beta_k$ is a nonzero (potentially complex-valued) function satisfying $\max_k |\beta_k|=1$.
Furthermore, it can be shown that the dissipative gap is given by 

\begin{equation}
\label{eq:dis:gap}
    \Delta_\mathcal{L} = \frac{\kappa}{2} \inf_{k} \vert \beta_k\vert^2.
\end{equation}
For fermionic density matrices, however, fermion-parity superselection~\cite{Wick1952aa,PhysRevA.104.032411} excludes the parity-odd coherence modes that realize the dissipative gap $\Delta_{\mathcal L}$ (see SM). The slowest physically accessible relaxation rate is therefore $\Gamma \equiv 2\Delta_{\mathcal L}$ and we characterize preparation by the asymptotic relaxation
timescale $\tau=\Gamma^{-1}$.

In this setting, we can rewrite the transformation matrices in~\cref{eq:bogoliubov:annihilator:from:real:space:operators} as $\mathbf{U}_{k\ell}=e^{ik\ell}u_k/\sqrt N$ and $\mathbf{V}_{k\ell}=e^{ik\ell}v_k/\sqrt N$ yielding the jump operator coefficients
\begin{equation}
\tilde{\mathbf{U}}_{j\ell}
=\frac{1}{N}\sum_k e^{ik(\ell-j)}u_k\beta_k,
\qquad
\tilde{\mathbf{V}}_{j\ell}
=\frac{1}{N}\sum_k e^{ik(\ell-j)}v_k\beta_k \, .
    \label{eq:U:tilde:V:tilde}
\end{equation}
These are the Fourier coefficients of the functions $u_k \beta_k$, $v_k \beta_k$, respectively. 
\Cref{eq:dis:gap,eq:U:tilde:V:tilde} thus directly connect the real-space locality of the jump operators to the preparation time through the momentum-space weights $\beta_k$. 
%
%
%

\begin{table}[]
    \centering
    \begin{tabular}{|c|c|c|c|}
    \hline
        Parameter regime & Phase & Jump operator support & $\tau$\\
        \hline
        \hline
        $|\mu|>2t$& trivial & $\mathcal O(e^{-r/\xi})$ & $\mathcal O(1)$ \\
        \hline
        $|\mu|=2t$, $\Delta>0$ & critical & $\mathcal O (r^{-n-1})$ & $\mathcal O(N^{2n})$\\
        \hline
        $|\mu|<2t$, $\Delta=0$ & critical & $\mathcal O (r^{-n-1})$ & $\mathcal O(N^{2n})$ \\
        \hline
        $|\mu|<2t$, $\Delta>0$ & topological & $\mathcal O (e^{-r/\xi})$ & $\mathcal O(1)$\\
        \hline
        $(\mu/2)^2 + \Delta^2 = t^2$ & topological & 2 & $\mathcal O(1)$
        \\
        \hline
        
    \end{tabular}
    \caption{Spatial decay of the jump operators and
scaling of the relaxation
timescale $\tau=\Gamma^{-1}$ across
the Kitaev-chain phase diagram. Here, $r$ denotes the distance from the reference site of the
jump operator and $\xi$ the localization length
of its coefficients, both measured in lattice spacings.}
    \label{tab:tradeoff}
\end{table}

\paragraph{Criticality versus topology--}
To determine how criticality and topology affect the
locality-preparation time tradeoff, we consider the Kitaev
chain~\cite{Kitaev2001}
\begin{equation}
    \hat H=\sum_{j=1}^{N}\Big[
    -t\,(\hat{c}_j^\dagger \hat{c}_{j+1}+\mathrm{h.c.})
    -\mu\Big(\hat{c}_j^\dagger \hat{c}_j-\frac12\Big)
    +\Delta\,(\hat{c}_j \hat{c}_{j+1}+\mathrm{h.c.})
    \Big],
\end{equation}
where $t$ is the nearest-neighbour hopping amplitude, $\mu$ is the chemical potential, and $\Delta$ is the $p$-wave pairing amplitude. For $|\mu|<2t$, the chain is in a topological superconducting phase, while for $|\mu|>2t$, the system is topologically trivial. 
These two gapped phases are separated by the critical lines at $|\mu|=2t$ and $\Delta>0$. 
The case $\Delta=0$ is special: the model reduces to a number-conserving tight-binding chain, which is gapless for $|\mu|<2t$, and trivially gapped otherwise.

Along the critical lines, the Bogoliubov coefficients develop discontinuities in momentum space, which inevitably lead to algebraically decaying real-space jump-operator coefficients. Their tails can only be suppressed by choosing the mixing function $\beta_k$ to vanish at the singular momentum $k^\star$. In this way, the regularity of both Fourier coefficients increases, changing the asymptotic behavior of the jump operators~\cite{nissila2018fourier} through~\cref{eq:U:tilde:V:tilde}. In particular, a zero of order $n$ smooths the products $\beta_k u_k$ and $\beta_k v_k$, yielding real-space coefficients that decay asymptotically as
$|j-l|^{-(n+1)}$. The same zero, however, necessarily closes the dissipative gap: since the momentum spacing scales as $1/N$, the smallest value sampled near $k^\star$ scales as $\inf_k |\beta_k|\sim N^{-n}$. \Cref{eq:dis:gap} then gives $\Gamma\sim \kappa N^{-2n}$, or equivalently a preparation time
$\tau\sim N^{2n}/\kappa$. Thus, progressively suppressing the spatial tails of the jump operators comes at the cost of an increasingly strong algebraic divergence of the preparation time, as illustrated in~\cref{fig:first}(b). From a physical perspective, propagating algebraic correlations in shorter time comes at the cost of an increased spatial range of the jump operators. We make these arguments rigorous in the End Matter, where we formally prove this tradeoff relation.
%
%

\begin{figure}
    \centering
    \begin{tikzpicture}
        \node[inner sep=0, anchor=south west] (img) at (0,0)
        {\includegraphics[width=\columnwidth]{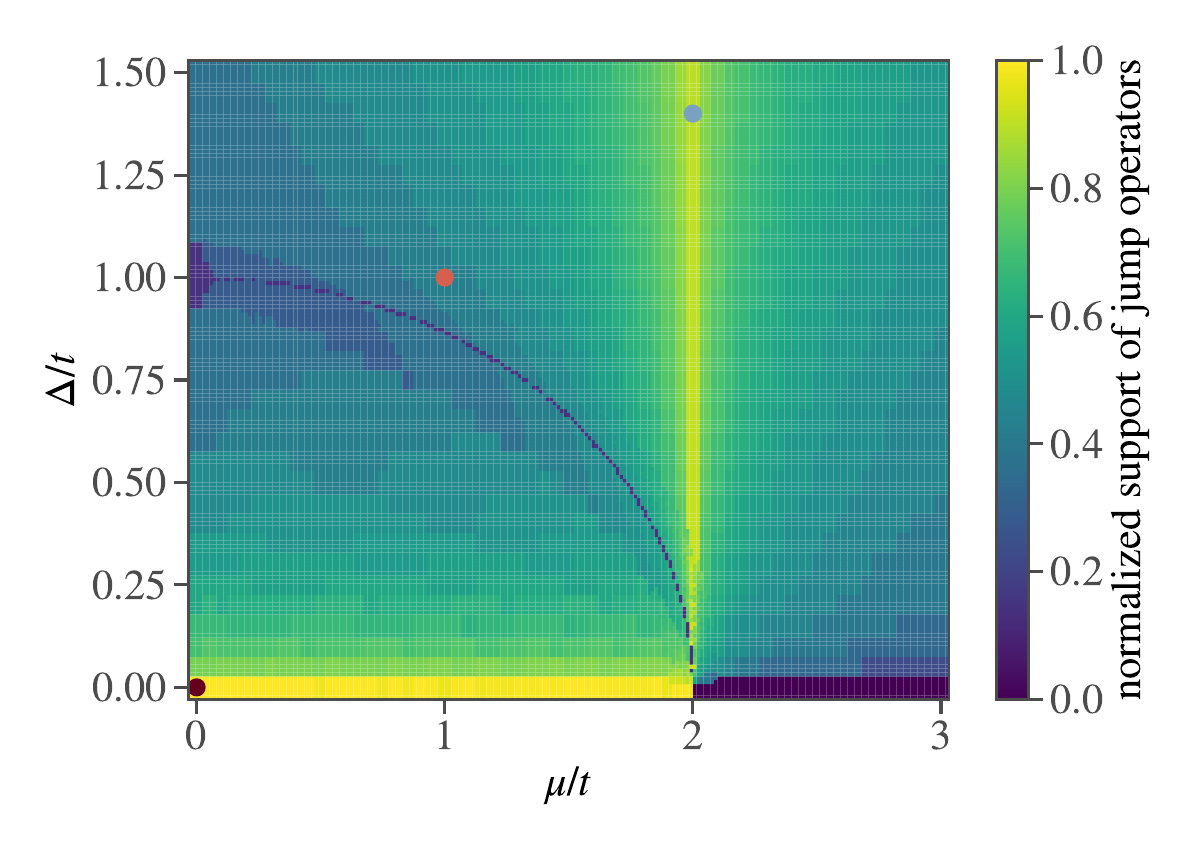}};

        \begin{scope}[x={(img.south east)}, y={(img.north west)}]

            \tikzset{
                phaselabel/.style={
                    font=\small,
                    fill=white,
                    fill opacity=0.75,
                    text opacity=1,
                    inner sep=1.5pt,
                    rounded corners=1pt
                },
                critlabel/.style={
                    font=\scriptsize,
                    fill=white,
                    fill opacity=0.75,
                    text opacity=1,
                    inner sep=1.2pt,
                    rounded corners=1pt
                }
            }

            \node[phaselabel] at (0.30,0.78) {topological};
            \node[phaselabel] at (0.72,0.80) {trivial};

            \node[critlabel] at (0.30,0.20) {XX criticality};
            \node[critlabel, rotate=90] at (0.60,0.62) {Ising criticality};

            \node[critlabel] at (0.31,0.48) {finite-range ellipse};
            \draw[->, thick]
            (0.3,0.51) -- (0.46,0.52);
        \end{scope}
    \end{tikzpicture}
    \caption{
    \textbf{Locality of the dissipator at fixed preparation time across the Kitaev phase diagram.} The support of the jump operators required to prepare the ground state at fixed dissipative gap $2\Delta_{\mathcal L}/\kappa=1$ is shown as a function of the chemical potential $\mu/t$ and pairing strength $\Delta/t$. The support is determined from the spatial extent of the real-space jump-operator coefficients above a cutoff $\epsilon=0.005$ for a system with $N=100$ sites.
    }
    \label{fig:phase:diagram}
\end{figure}
In contrast, in the topologically trivial gapped phase, the Bogoliubov coefficients can be chosen to be analytic in momentum space, yielding exponentially localized jump operators. Since no suppression of momentum modes is required, $\beta_k$ can remain bounded away from zero, so the dissipative gap stays finite and $\tau=\mathcal{O}(1)$ in the thermodynamic limit.

The topological phase is similar: although nontrivial winding occurs, via a suitable gauge we can construct exponentially localized jump operators without closing the dissipative gap, for an arbitrary smooth nonzero mixing function $\beta_k$.  
Thus, topology alone does not impose a locality-preparation time tradeoff.
Finally, along the special ellipse
$\Delta^2+(\mu/2)^2=t^2$ in the topological regime, by choosing a specific $\beta_k$ the optimized jump operators become strictly finite range
\begin{equation}
    \hat L_j = \sqrt{\kappa}\left( 
    -\Delta \hat c_j
    + \frac{2t+\mu}{2}\hat c_j^\dagger
    + \Delta \hat c_{j+1}
    + \frac{2t+\mu}{2}\hat c_{j+1}^\dagger \right) .
    \label{eq:jump:ops:on:ellipse}
\end{equation}
This reflects the fact that along this special ellipse $\Delta^2+(\mu/2)^2=t^2$, the target state can be written as a superposition of two product states.
At $\mu=0$ and $\Delta=t$,~\cref{eq:jump:ops:on:ellipse} reduces to the dissipative construction of Ref.~\cite{Diehl_2011}.
The locality-preparation time scaling across the Kitaev phase diagram is summarized in \cref{tab:tradeoff}. 

To visualize the same constraints at a fixed preparation timescale, in \cref{fig:phase:diagram} we set the dissipative gap to $2\Delta_{\mathcal L}/\kappa=1$ (i.e. choosing $n=0$ along the critical lines) and show the normalized support of the corresponding jump operators across the phase diagram. The horizontal and vertical bright lines mark the critical regions, where the jump operators extend over the entire system, whereas in the gapped phases they remain localized. 

\begin{figure}
    \centering
    \includegraphics[width=\columnwidth]{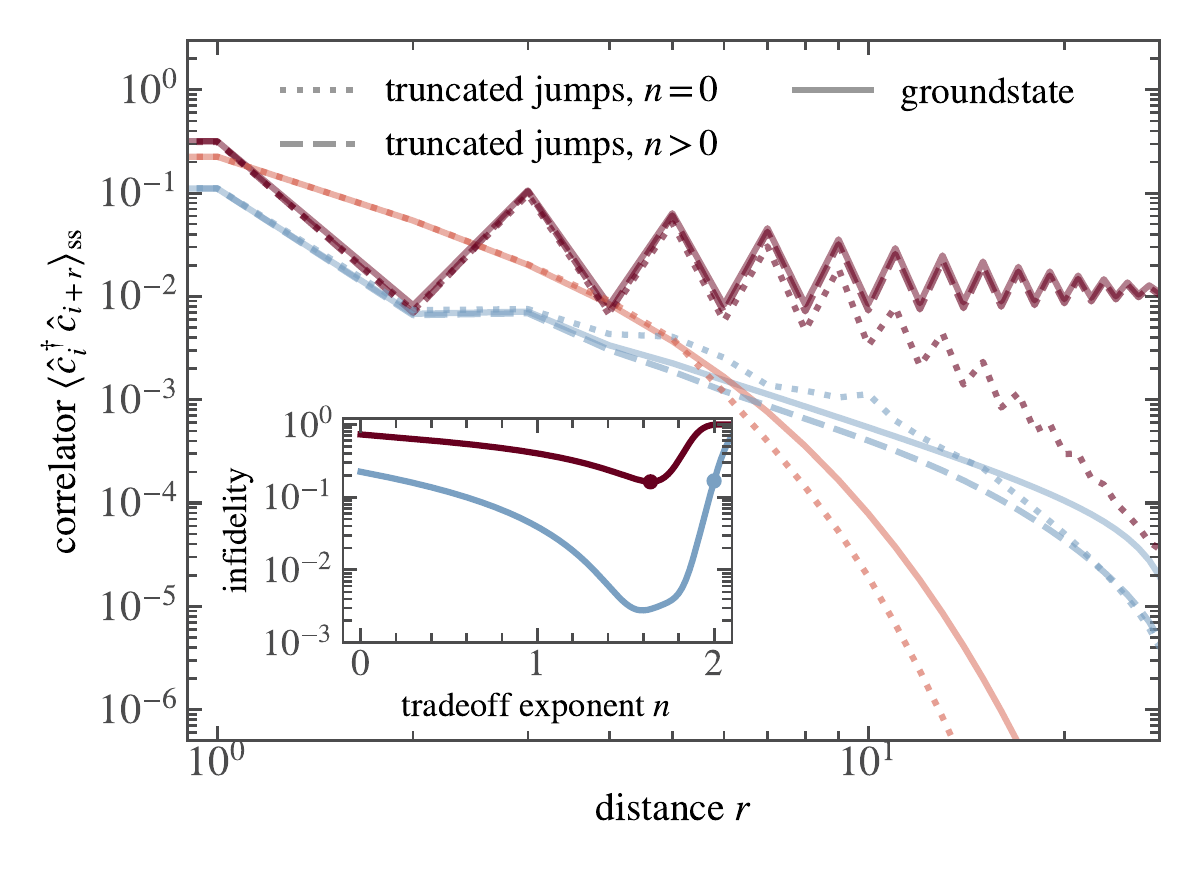}
\caption{\textbf{Accuracy of finite-range dissipators.}
Steady-state correlator $|\langle \hat c_i^\dagger \hat c_{i+r}\rangle_{\rm ss}|$ obtained after truncating the jump operators, compared with the exact ground-state result (solid lines). Dotted lines correspond to direct truncation of the real-space Bogoliubov annihilators, i.e., the trivial mixing $\beta_k=1$ ($n=0$). Dashed lines instead exploit a nontrivial mixing $\beta_k$ ($n>0$) to suppress the momentum-space discontinuities and improve the spatial decay of the jump operators before applying the same finite-range truncation. The inset shows the steady-state infidelity $1-\mathcal{F}_{\rm HS}$, where
$\mathcal{F}_{\rm HS}$ is the normalized Hilbert-Schmidt overlap, as a function of the suppression exponent $n$ for the two critical points. Highlighted points indicate the values of $n$ used for the dashed curves in the main panel. Results are shown for $N=61$ sites and jump operators truncated to $|j-l|<5$, at the XX critical point $(\mu/t,\Delta/t)=(0,0)$ with $n=1.65$ (dark red), the Ising critical point $(2,1.4)$ with $n=2$ (blue), and the topological point $(1,1)$ (orange), highlighted by the dots with the corresponding colors in~\cref{fig:phase:diagram}.
}
    \label{fig:steady:state}
\end{figure}

In practice, however, the available dissipative couplings will generally be restricted to a finite spatial range. It is therefore crucial to determine how truncating the tails of the jump operators affects the accuracy of the prepared steady state, both at the level of physical observables and of the global state. In \cref{fig:steady:state}, we impose the same finite-range cutoff on the three representative points highlighted in \cref{fig:phase:diagram} and compare the resulting steady-state correlations with those of the exact target ground state. At the critical points, directly truncating the real-space Bogoliubov annihilators ($n=0$) produces sizable deviations. By contrast, for $n>0$ we first exploit the nontrivial mixing
$\beta_k=|\cos k|^n$ at XX criticality and
$\beta_k=|\cos(k/2)|^n$ at Ising criticality,
which suppress the momentum-space singularities and thereby the algebraic real-space tails before applying the same truncation. This substantially improves the agreement at fixed spatial range. In the gapped topological phase, where the jump operators are exponentially localized, direct truncation remains more accurate. The inset quantifies the corresponding global error through the steady-state infidelity, showing that an appropriate choice of the suppression exponent $n$ can strongly improve the fidelity of the finite-range dissipator. 
Note that the fidelity curve shows a non-monotonicity with the tradeoff exponent, which originates from the interplay between real-space truncation and strong suppression of the momentum prefactors close to the point of discontinuity. While the jump operator coefficients close to the momentum discontinuity are very small for large $n$, the error through truncation can remain comparably large. Since for the fidelity the ratio of prefactors matters, the truncation errors can get strongly amplified (see SM). 

\paragraph{Discussion and outlook--}
We have developed a general framework for optimizing the locality of dissipative ground-state preparation in quadratic fermionic systems, exploiting the freedom to mix Bogoliubov annihilation operators without changing the target state. The resulting exact locality-preparation time relation is controlled by the momentum-space regularity of the Bogoliubov coefficients: at criticality, the more strongly the algebraic tails of the jump operators are suppressed, the more severely the preparation time grows (polynomially) with system size. By contrast, both trivial and topological gapped phases admit exponentially localized jump operators while retaining a finite preparation time in the thermodynamic limit. These results establish dissipator locality as a quantitative resource for scalable DSP, linking the required interaction range directly to the achievable preparation time.

The odd- and even-parity character of the low-lying Liouvillian modes also leads to an interesting initial-state dependence of the relaxation dynamics.
While fermion-parity superselection excludes the slowest mode for physical fermionic states, these modes become accessible in spin realizations.
As shown in the SM, eliminating their initial overlap yields a twofold Mpemba-type preparation speedup~\cite{Carollo2021,ares2025quantum, Westhoff2025}.

Looking ahead, an important direction is to investigate whether measurements, feedback, and adaptive open-system dynamics~\cite{Pocklington2025adaptive} can mitigate or circumvent the locality-preparation time tradeoffs derived here. Moreover, it would be valuable to explore how these relations are modified by non-Markovian and spectrally structured reservoirs~\cite{p8n2-jzgh}, where memory effects and frequency selectivity provide additional resources for state preparation. Extending the analysis beyond quadratic systems represents a further central challenge. Interacting many-body models could be studied numerically using methods such as CLIK-MPS~\cite{westhoff2026tensornetworkframework}, which enables the computation of dissipative gaps in large-scale systems. Finally, translating the optimized dissipators into experimental protocols will require accounting for restrictions on available couplings and reservoir engineering in platforms such as ultracold atoms~\cite{PhysRevLett.111.185301, Zeiher2016-ht}, trapped ions~\cite{RevModPhys.93.025001}, and superconducting-qubit processors~\cite{Arute2019-bd}.

\paragraph{Note added} After the first version of this manuscript appeared on arXiv, a related work by Tang, Barad, and Wen~\cite{tang2026localitydissipativegroundstatepreparation} appeared, studying criteria for the existence of exponentially local linear jump operators for dissipative ground-state preparation in free-fermion systems.

\paragraph{Acknowledgements} 
We thank Artur M. Lacerda and John Goold for enlightening discussions about dissipative fermionic systems.
PW acknowledges support by the Deutsche Forschungsgemeinschaft (DFG, German Research Foundation) under Germany’s Excellence Strategy-426 EXC-2111-390814868.
MM acknowledges funding from the European Union’s Horizon research and innovation programme under the Marie Skłodowska-Curie grant agreement No 101264565.

\FloatBarrier
\bibliographystyle{apsrev4-2}
\bibliography{Literature}
\FloatBarrier

\section*{End matter}
Here we provide a rigorous proof of the locality-preparation time tradeoff relation that we sketched in the main text.
We consider the jump operator coefficients \cref{eq:mixing:matrix}, which fix the jump operators through \cref{eq:jump:operators:real} and (\ref{eq:U:tilde:V:tilde}).
Thus the locality of the jump operators is governed by the Fourier coefficients of the two functions
\begin{equation}
    f_u(k)=\beta(k)\,u(k),\qquad f_v(k)=\beta(k)\,v(k),
    \label{eq:fu_fv}
\end{equation}
where we write the dependence on $k$ now explicitly.
\paragraph{Regularity and Fourier decay.}
For $n\in\mathbb{N}_0$ and $\alpha\in(0,1]$, a $2\pi$-periodic function $f$
belongs to the H\"older class $C^{n,\alpha}$ if it is $n$ times continuously
differentiable and its $n$-th derivative is uniformly $\alpha$-H\"older
continuous, $|f^{(n)}(k+h)-f^{(n)}(k)|\le C|h|^{\alpha}$. We use the sharp
Fourier-decay estimate of Ref.~\cite{nissila2018fourier}. In the direction from
regularity to decay, two further hypotheses are required: $f^{(n)}$ must be
absolutely continuous, and the number of local extrema of
$\Delta_h f^{(n)}(k):=f^{(n)}(k+h)-f^{(n)}(k)$ must stay bounded for small $h$.
Under these conditions,
\begin{equation}
    f\in C^{n,\alpha}
    \;\Longrightarrow\;
    \hat f_r=O\!\left(|r|^{-(n+\alpha+1)}\right),
    \label{eq:holder_to_fourier}
\end{equation}
where $\hat f_r$ denotes the $r$-th Fourier coefficient of $f$.
In our setting $f_u$ and $f_v$ are piecewise smooth
with a finite number of isolated algebraic singularities.
$\alpha=1$ (Lipschitz) is included, producing integer
tail exponents.
Also the converse, softer statement holds
\begin{equation}
    \hat f_r=O(|r|^{-s-1}) 
    \;\Longrightarrow\;
    f\in C^{n',\alpha'} \;\; \forall n', \alpha': n'+\alpha'<s .
    \label{eq:fourier_to_holder}
\end{equation}
Consequently, tails decaying faster than the
jump rate $|r|^{-1}$ force $f_u,f_v$ to be continuous.
\vspace{0.2cm}
\paragraph{Critical discontinuity.}
At the Ising critical points the Bogoliubov coefficients have, near the
singular momentum $x=\pi$, the structure
\begin{equation}
    u(k)=\operatorname{sgn}(k-x)\,u_0(k),\qquad v(k)=v_0(k),
    \label{eq:sign_jump_general}
\end{equation}
with $u_0,v_0$ regular across $x$ and $u_0(x),v_0(x)\neq0$
(cf. Supplemental material).
In close analogy, on the XX critical line, the coefficients again have a discontinuity
\begin{equation}
    u(k) = \Theta(k_0-|k|), \qquad v(k) = 1 - u(k),
    \label{eq:sign_jump_xx}
\end{equation}
for some $k_0$.
\vspace{0.2cm}
\paragraph{Lemma.}
Assume the local structure~\eqref{eq:sign_jump_general} or \eqref{eq:sign_jump_xx}. Then $f_u=\beta u$ and
$f_v=\beta v$ both belong to $C^{n,\alpha}$ across $x$ \textit{if and only if}
$\beta\in C^{n,\alpha}$ across $x$ with
\begin{equation}
    \beta^{(m)}(x)=0,\qquad m=0,\ldots,n.
    \label{eq:beta_derivatives_zero}
\end{equation}
\textit{Proof.}
For the forward direction and the local structure \cref{eq:sign_jump_general}, since $v_0(x)\neq0$ and $f_v=\beta v_0$ is regular across $x$, we may write
$\beta=f_v/v_0$ locally, so $\beta$ inherits the $C^{n,\alpha}$ regularity of
$f_v$. Set $g(k)=\beta(k)u_0(k)$, so that
$f_u(k)=\operatorname{sgn}(k-x)\,g(k)$. Continuity of $f_u$ at $x$ requires its
one-sided limits to agree; as the sign flips, this forces $g(x)=0$, and hence
$\beta(x)=0$ since $u_0(x)\neq0$. Proceeding inductively, suppose
$\beta^{(j)}(x)=0$ for all $j<n$. Away from $x$ the sign is constant, so
continuity of the $n$-th derivative of $f_u$ requires $g^{(n)}(x)=0$; by Leibniz,
\begin{equation}
    g^{(n)}(x)=\sum_{j=0}^{n}\binom{n}{j}\beta^{(j)}(x)\,u_0^{(n-j)}(x)
    =\beta^{(n)}(x)\,u_0(x),
\end{equation}
so $\beta^{(n)}(x)=0$. 
Conversely, if $\beta\in C^{n,\alpha}$ with
Eq.~\eqref{eq:beta_derivatives_zero}, the vanishing derivatives cancel the sign
discontinuity through order $n$, and the $\alpha$-H\"older continuity of
$\beta^{(n)}$ gives $f_u,f_v\in C^{n,\alpha}$.
For the structure \eqref{eq:sign_jump_xx}, the arguments are analogous. The regularity of $\beta$ directly follows from $f_u+f_v=\beta$, continuity of $f_u$ requires the one-sided limits at $x=\pm k_0$ to agree, which directly implies $\beta(x)=0$ since $f_u(k)= 0 \;\; \forall |k|>|x|$.
Also, the induction can be carried out similarly as $u$ itself is constant away from $x$.
The converse direction is identical to the one above.\hfill$\square$
\vspace{0.2cm}
\paragraph{Tradeoff.}
Combining the Lemma with Eq.~\eqref{eq:holder_to_fourier}, a dissipative gap
$\Delta_\mathcal L\sim N^{-2(n+\alpha)}$ implies at best a Fourier decay of
\begin{equation}
    \max\!\left(|\tilde u_r|,|\tilde v_r|\right)=O\!\left(|r|^{-(n+\alpha+1)}\right),
    \label{eq:jump_tail_bound}
\end{equation}
up to subpolynomial corrections.

If on the other hand \cref{eq:jump_tail_bound} holds, the gap at least closes at a rate $\Delta_\mathcal L \sim N^{-2(n+\alpha)}$ up to subpolynomial corrections.

\vspace{0.3cm}

Already at $n=0$ the obstruction appears: by the Lemma, making $f_u$ and $f_v$
merely continuous across $x$ requires $\beta(x)=0$, so $\beta$ cannot stay
bounded away from zero. 
More generally, the conditions on the dissipative gap dictate that the highest order zero crossing of $\beta$ is of order $n+\alpha$. This directly implies that if $\alpha\neq0$ at most the first $n$ derivatives of $\beta$ can vanish at $x$, and $\beta$ can at most be in $C^{n, \alpha}$, if $n+\alpha$ is not a whole number. Assume first that $\beta$ is not $C^{n,\alpha}$ or \cref{eq:beta_derivatives_zero} is not fulfilled. 
Then at least one of $f_u$ and $f_v$ is also not $C^{n, \alpha}$ (w.l.o.g. assume it is $f_u$) by the Lemma and by \cref{eq:fourier_to_holder} $\tilde u_r$ cannot decay like $|r|^{-s-1}$ for some $s<n+\alpha$. 
On the other hand, if $\beta$ is $C^{n,\alpha}$ and fulfills \cref{eq:beta_derivatives_zero}, then $f_u$ and $f_v$ are $C^{n, \alpha}$ and by \cref{eq:holder_to_fourier} we get the desired decay. 
In the special case of an integer exponent, then, if $\beta$ is in $C^{n, 0}$ it is also in any $C^{n-1, \alpha'}$ for $\alpha'\leq 1$, leading to the desired scaling through the same argumentation.
For the converse direction, \cref{eq:jump_tail_bound} implies that $f_u$ and $f_v$ are in $C^{n', \alpha'}$ with $n'+\alpha'<n+\alpha$. By the Lemma III $\beta$ obeys Eq.~\eqref{eq:beta_derivatives_zero} together with corresponding $\alpha$-H\"older continuity.
On a chain of $N$ sites the momenta are spaced by $2\pi/N$,
so, provided $x$ is not itself a sampled momentum, the smallest sampled value of
$|\beta_k|$ near $x$ scales as $\inf_k|\beta_k|\sim N^{-(n'+\alpha')}$, by the Hölder continuity.
Consequently, we obtain that the gap closes faster than $N^{-2a}$ for any $a<n+\alpha$. \hfill$\square$
This is the locality-preparation time tradeoff: shortening the tails from
$|r|^{-1}$ to $|r|^{-(n+\alpha+1)}$ requires a zero of order $n+\alpha$ in
$\beta$, which drives the preparation time up as $N^{2(n+\alpha)}$. The main-text
constructions $\beta_k=|\cos(k/2)|^{n}$ (Ising) and $\beta_k=|\cos k|^{n}$ (XX)
have a zero of integer order $n$, i.e.\ the Lipschitz case $\alpha=1$, giving
$|r|^{-(n+1)}$ and $\Delta_{\mathcal L}\sim N^{-2n}$.
\end{document}


\def\thetitle{Supplemental Material for \\``Locality-Preparation Time Tradeoffs in Dissipative Quantum Systems"} 
\title{\thetitle}
\author{Philipp Westhoff}\email{p.westhoff@physik.uni-muenchen.de}
\affiliation{\affilA}
\author{Sara Murciano}\email{sara.murciano@universite-paris-saclay.fr}
\affiliation{\affilB}
\author{Mattia Moroder}
\email{moroderm@tcd.ie}
\affiliation{\affilC}
\affiliation{\affilD}
\maketitle

\appendix
\onecolumngrid
\section{General Form of the Jump Operators}
Consider a fermionic Hamiltonian diagonalized in a Bogoliubov basis
\begin{equation}
    \hat H = \sum_k \varepsilon_k\, \hat a_k^\dagger \hat a_k,
\end{equation}
where $\hat a_k$ are fermionic annihilation operators.
Note that we may restrict all our analysis to the case $\varepsilon_k \geq 0$, since a general quadratic fermionic Hamiltonian can always be brought into this form by a particle-hole transformation~\cite{Kitaev2001}.
This covers a wide range of physical systems, including, for instance, the Su-Schrieffer-Heeger model, the Kitaev chain and the transverse-field Ising model.
The ground state in this Bogoliubov eigenbasis is simply the vacuum $\ket{\Omega}$ (i.e. $\hat a_k \ket{\Omega} = 0  \hspace{0.3cm}\forall k$).
Thus, if we choose the simple jump operator 
\begin{equation}
    \hat{L}_k = \sqrt{\kappa}\, \hat{a}_k
    \label{eq:app:simple:jump:op}
\end{equation}
for each mode, the Lindbladian~\cite{Lindblad1976b}
\begin{equation}
 \frac{d}{dt} \hat\rho = -i [\hat H, \hat\rho] + \sum_k \mathcal{D}_{\hat L_k}[\hat\rho] \equiv \mathcal{L}[\hat \rho]
 \label{eq:lindblad}
\end{equation}
will drive any initial state to the ground state. 
Here $\hat\rho$ is the density matrix of the system, $\mathcal{D}_{\hat L_k}[\hat\rho] = \hat L_k \hat\rho \hat L_k^\dagger - \frac{1}{2}\{\hat L_k^\dagger \hat L_k, \hat\rho\}$ is the dissipator associated with jump operator $\hat L_k$ and $\mathcal{L}$ is the Lindbladian superoperator.
With~\cref{eq:app:simple:jump:op}, each dissipator describes 
amplitude damping of mode $k$ at rate $\kappa$.
Note that in the case of zero modes ($k'$ s.t. $\epsilon_{k'}=0$), the ground state is degenerate and the choice of jump operators depends on the specific ground state which is of interest. Depending on that, the corresponding jump operator needs to be chosen as annihilation or creation.

In the following, we will prove that a much richer family of jump operators than~\cref{eq:app:simple:jump:op} drives the system to the target Bogoliubov vacuum state $\ket{\Omega}$ if we only consider dissipative dynamics and omit the Hamiltonian contribution. These must satisfy~\cite{Kraus2008}:
\begin{enumerate}
\item $\forall k:\ \hat L_k\ket{\Omega}=0$
\quad ($\ket{\Omega}$ is a dark state of each $\hat L_k$),
\item $\forall \ket{\psi}\neq \ket{\Omega}\ \exists k:\ \hat L_k\ket{\psi}\neq 0$
\quad ($\ket{\Omega}$ is the unique dark state).
\end{enumerate}
We consider jump operators which are linear in the creation and annihilation operators. Then, property 1. directly implies that
\begin{equation}
    \hat L_k = \sum_{\ell} \mathbf B_{k\ell}\, \hat a_\ell,
    \label{eq:general:jump:op}
\end{equation}
i.e.\ each $\hat L_k$ is a linear combination of quasiparticle annihilation operators $\hat a_\ell$.
In fact, these conditions on $\hat L_k$ are equivalent to:
\begin{equation}
    \ket{\Omega}\ \text{is a unique dark state}
\quad \Longleftrightarrow \quad
\hat L_k=\sum_\ell \mathbf B_{k\ell}\hat a_\ell
\ \text{with}\ \mathbf{B}\ \text{of full rank}.
\end{equation}
We call $\mathbf{B}$ the mixing matrix.

\vspace{0.2cm}

\paragraph*{Proof (\texorpdfstring{$\Rightarrow$}{⇒} necessity of full rank)}
Assume that $\mathbf{B}$ does \emph{not} have full rank. Then there exists a nonzero vector
$\mathbf{v}\neq 0$ with $\|\mathbf{v}\|=1$ such that
\begin{equation}
    \mathbf{B} \mathbf{v} = 0.
\end{equation}
Define the (one-quasiparticle) state
\begin{equation}
    \ket{\psi} = \sum_{i=1}^{N} \mathbf{v}_i\, \hat a_i^\dagger \ket{\Omega}.
\end{equation}

Then for every $k$,
\begin{align}
    \hat L_k\ket{\psi}
    &=
    \sum_\ell \mathbf B_{k\ell}\hat a_\ell
    \sum_i \mathbf{v}_i \hat a_i^\dagger \ket{\Omega}
    =
    \sum_{\ell,i} \mathbf B_{k\ell} \mathbf{v}_i\, \hat a_\ell \hat a_i^\dagger \ket{\Omega}
    \nonumber\\
    &=
    \sum_{\ell,i} \mathbf B_{k\ell} \mathbf{v}_i\, \delta_{\ell i}\ket{\Omega}
    =
    \left(\sum_i \mathbf B_{k i} \mathbf{v}_i\right)\ket{\Omega}
    =
    (\mathbf B\mathbf{v})_k \ket{\Omega}
    =
    0.
\end{align}
Hence $\ket{\psi}\neq\ket{\Omega}$ is also a dark state, contradicting uniqueness.

\vspace{0.2cm}

\paragraph*{Proof (\texorpdfstring{$\Leftarrow$}{<=} sufficiency of full rank)}  

Assume $\mathbf{B}$ has full rank. Let $\ket{\psi}\neq \ket{\Omega}$.
Then there exists at least one mode index $\ell$ such that
\begin{equation}
    \hat a_\ell \ket{\psi}\neq 0.
\end{equation}
Define the vector
\begin{equation}
    A(\psi) =
    \big(
    \hat a_1\ket{\psi},
    \hat a_2\ket{\psi},
    \ldots,
    \hat a_N\ket{\psi}
    \big)^{T}
    \neq 0.
\end{equation}
Now
\begin{equation}
    \big(
    \hat L_1\ket{\psi},
    \ldots,
    \hat L_N\ket{\psi}
    \big)^{T}
    =
    \mathbf{B}\, A(\psi).
\end{equation}
Since $\mathbf{B}$ is full rank and $A(\psi)\neq 0$, we have $\mathbf{B}\,A(\psi)\neq 0$.
Therefore, there exists some $k$ such that $\hat L_k\ket{\psi}\neq 0$.
This proves the uniqueness of the dark state.

\section{Spectral decomposition of the Lindbladian}
\label{app:sec:spectral}
\subsection{Spectrum for decoupled quasiparticle modes}
\label{subsec:spectrum:decoupled}
We first consider the diagonal choice of jump operators
\begin{equation}
    \hat L_k = \sqrt{\kappa_k}\,\hat a_k ,
\end{equation}
for which the Lindbladian decomposes into independent quasiparticle modes,
\begin{equation}
    \mathcal L = \sum_{k=1}^N \mathcal L^{(k)} .
\end{equation}
For a single mode $k$, the four Lindbladian eigenvalues are
\begin{equation}
    \lambda_0^{(k)}=0,\qquad
    \lambda_{\pm}^{(k)}
    =-\frac{\kappa_k}{2}\pm i\varepsilon_k,\qquad
    \lambda_2^{(k)}=-\kappa_k .
\end{equation}
The modes with eigenvalues $\lambda_{\pm}^{(k)}$ correspond to coherences
between the vacuum and the single-quasiparticle state, whereas
$\lambda_2^{(k)}$ describes population relaxation.
Since the different quasiparticle sectors are independent, the full
Lindbladian spectrum is obtained by summing the corresponding single-mode
eigenvalues. Introducing
$\bm{\eta}=(\eta_1,\ldots,\eta_N)$ with
$\eta_k\in\{0,+,-,2\}$, one has
\begin{equation}
    \lambda_{\bm{\eta}}
    =
    \sum_{k=1}^N \lambda_{\eta_k}^{(k)} .
\end{equation}
Consequently,
\begin{equation}
    \operatorname{Re}\lambda_{\bm{\eta}}
    =
    -\frac12\sum_k \kappa_k |\eta_k|,
\end{equation}
where $|\eta_k|=0,1,1,2$ for
$\eta_k=0,+,-,2$, respectively.
Let
\begin{equation}
    k_0=\operatorname*{arg\,min}_k \kappa_k .
\end{equation}
The slowest nonstationary Lindbladian
modes have eigenvalues
\begin{equation}
    \lambda_{\pm}^{(k_0)}
    =
    -\frac{\kappa_{k_0}}{2}
    \pm i\varepsilon_{k_0},
\end{equation}
and hence the spectral gap of the unrestricted Lindbladian is
\begin{equation}
    \Delta_{\mathcal L}
    =\frac12\min_k\kappa_k .
\end{equation}
These slow modes correspond to coherences between the vacuum and single-quasiparticle states. For fermionic systems, however, such coherences connect states of opposite fermion parity and are excluded by the fermion-parity superselection rule. Consequently, the slowest decay rate accessible to physical fermionic density matrices is
\begin{equation}
    \Gamma = \min_k \kappa_k ,
\end{equation}
rather than the unrestricted Lindbladian gap $\Delta_{\mathcal L}=\frac{1}{2}\min_k\kappa_k$. In spin or hard-core-boson systems, where no analogous parity superselection rule applies, vacuum--single-particle coherences are physical and relaxation at the slower rate $\frac{1}{2}\min_k\kappa_k$ can occur.

\subsection{Diagonalization for the general jump operators}
Here we study the most general jump operators~\cref{eq:general:jump:op}.
Since $\mathbf{B}$ mixes the $k$ modes, the Lindbladian no longer factorizes. Nevertheless, it remains quadratic in the fermionic operators, and third quantization~\cite{Prosen_2008} can be used to extract its full spectrum.

We first transform to a Majorana basis by defining $2N$ Majorana operators
\begin{equation}
\hat w_{2j-1}=\hat a_j+\hat a_j^\dagger,
\qquad
\hat w_{2j}=i(\hat a_j-\hat a_j^\dagger),
\qquad j=1,\ldots,N.
\end{equation}
The Hamiltonian then takes the form
\begin{equation}
\hat H=
\frac{i}{4}\sum_j \varepsilon_j
\left(
\hat w_{2j}\hat w_{2j-1}
-
\hat w_{2j-1}\hat w_{2j}
\right)
=: \sum_{\mu\nu}\hat w_\mu h_{\mu\nu}\hat w_\nu ,
\end{equation}
which defines the antisymmetric $2N\times 2N$ matrix $h$, whose elements can be read off directly. Similarly, the jump operators read
\begin{equation}
    \hat L_k
=
\frac{1}{2}\sum_j \mathbf B_{kj}
\left(
\hat w_{2j-1}-i\hat w_{2j}
\right)
=: \sum_m l_{km}\hat w_m ,
\end{equation}
introducing the $N\times 2N$ matrix $l$ whose elements are likewise read off directly. We further define the $2N\times 2N$ matrix
\begin{equation}
    M_{nm} = \sum_\mu l_{\mu n}\,l^*_{\mu m},
\end{equation}
and the $4N\times 4N$ matrix $\mathbf{A}$ with elements
\begin{align}
    \mathbf{A}_{2n-1,\,2m-1} &= -2i\,h_{nm} - M_{nm} + M_{mn}, \\
    \mathbf{A}_{2n-1,\,2m}   &=  2i\,M_{mn}, \\
    \mathbf{A}_{2n,\,2m-1}   &= -2i\,M_{nm}, \\
    \mathbf{A}_{2n,\,2m}     &= -2i\,h_{nm} + M_{nm} - M_{mn}.
\end{align}
The eigenvalues of the Lindbladian are obtained by diagonalizing $\mathbf{A}$,
which yields eigenvalues $\zeta_j$. Of the $4N$ eigenvalues of $\mathbf{A}$,
exactly $2N$ have positive real part; only these enter the Lindbladian spectrum.
The Lindbladian eigenvalues are labeled by a binary vector
$\bm{\nu}\in\{0,1\}^{2N}$ and read
\begin{equation}
    \lambda_{\bm{\nu}} = -2\sum_j \zeta_j\,\nu_j.
\end{equation}
The corresponding eigenmatrices can be constructed from the eigenvectors
of $\mathbf{A}$.

Two remarks are in order. First, the dissipative gap is completely determined
by the smallest (in real part) eigenvalue of $\mathbf{A}$, so the spectral
properties of this matrix give complete information about the gap scaling of
the Lindbladian. Second, although $\mathbf{A}$ may not be analytically
diagonalizable for large systems and arbitrary $\mathbf{B}$, third
quantization reduces the complexity of the problem from exponential to
polynomial in $N$.

\subsection{Special case: unitary mixing matrix and generalizations}
\label{app:subsec:unitary:B}

For a general mixing matrix $\mathbf{B}$, extracting the spectrum requires
third quantization as described above. In many applications, however,
symmetry considerations impose further structure on $\mathbf{B}$.
Here we consider the family of matrices of the form
\begin{equation}
\label{eq:more:general:B}
    \mathbf B_{jk} = \mathbf{U}_{jk}\beta_k,
\end{equation}
where $\mathbf{U}$ is unitary and $\beta_k \neq 0$, so that $\mathbf{B}$ has full rank.
This turns out to be a remarkably powerful class: as we now show, the
Lindbladian spectrum is identical to that of the simple diagonal case.

Let $\hat{\bm a}$ denote the vector of Bogoliubov annihilation operators,
with $(\hat{\bm a})_k=\hat a_k$, and consider a general dispersion
$\varepsilon_k\geq 0$. The Hamiltonian part of the Lindbladian is already
diagonal, so we focus on the dissipative part. The jump operators read
\begin{equation}
    \hat L_j=\sum_k \mathbf B_{jk}\hat a_k=[\mathbf B\hat{\bm a}]_j .
\end{equation}
The corresponding dissipators are
\begin{align}
\mathcal D_j[\hat\rho]
&=
\hat L_j\hat\rho\hat L_j^\dagger
-\frac{1}{2}\hat L_j^\dagger\hat L_j\hat\rho
-\frac{1}{2}\hat\rho\hat L_j^\dagger\hat L_j
\nonumber\\
&=
\sum_{km}\mathbf B_{jm}^* \mathbf B_{jk}
\left(
\hat a_k\hat\rho\hat a_m^\dagger
-\frac{1}{2}\hat a_m^\dagger\hat a_k\hat\rho
-\frac{1}{2}\hat\rho\hat a_m^\dagger\hat a_k
\right).
\end{align}
Defining
\begin{equation}
\mathcal D_{km}[\hat\rho]
:=
\hat a_k\hat\rho\hat a_m^\dagger
-\frac{1}{2}\hat a_m^\dagger\hat a_k\hat\rho
-\frac{1}{2}\hat\rho\hat a_m^\dagger\hat a_k ,
\end{equation}
and summing over all jump operators $j$, the total dissipator becomes
\begin{equation}
\mathcal D[\hat\rho]
=
\sum_j\mathcal D_j[\hat\rho]
=
\sum_{km}\mathcal D_{km}[\hat\rho]
\sum_j \mathbf B_{jm}^* \mathbf B_{jk}
=
\sum_{km}\mathcal D_{km}[\hat\rho]\,[\mathbf B^\dagger \mathbf B]_{mk}.
\label{eq:app:dissipator:with:B}
\end{equation}
For the specific choice~\cref{eq:more:general:B}, unitarity of $\mathbf{U}$ gives
\begin{equation}
[\mathbf B^\dagger \mathbf B]_{mk}
=
\beta_m^*[U^\dagger U]_{mk}\beta_k
=
\delta_{mk}|\beta_k|^2 .
\end{equation}
Hence~\cref{eq:app:dissipator:with:B} reduces to
\begin{equation}
\mathcal D[\hat\rho]
=
\sum_m |\beta_m|^2\,\mathcal D_{mm}[\hat\rho]
=
\sum_m |\beta_m|^2
\left(
\hat a_m\hat\rho\hat a_m^\dagger
-\frac{1}{2}\hat a_m^\dagger\hat a_m\hat\rho
-\frac{1}{2}\hat\rho\hat a_m^\dagger\hat a_m
\right),
\end{equation}
which is precisely the simple decoupled dissipator with rates $|\beta_m|^2$.
Hence the spectrum and eigenmatrices are unchanged with respect to the
diagonal case, and the results of~\cref{subsec:spectrum:decoupled} apply
directly. In particular, the dissipative gap reads
\begin{equation}
\label{eq:dissipative:gap}
    \Delta_{\mathcal{L}}
    := \inf_{n:\,\mathrm{Re}\,\lambda_n < 0}\lvert\mathrm{Re}\,\lambda_n\rvert
    = \frac{1}{2}\inf_k\,|\beta_k|^2 \, .
\end{equation}

As discussed in~\cref{subsec:spectrum:decoupled}, for physical fermionic density matrices the parity-odd coherence modes associated with this gap are excluded by fermion-parity superselection. The slowest physically accessible decay rate is therefore $2\Delta_{\mathcal L}$.

\subsection{Restrictions on the mixing matrix for periodic systems}
\label{app:subsec:periodic:systems}
For systems with periodic boundary conditions, physical considerations
impose additional constraints on the jump operators: they should respect
the translation invariance of the system, meaning they transform covariantly
under translations. Denoting by $\mathcal{T}$ the translation operator
acting as $\mathcal{T}\hat{c}_j = \hat{c}_{j+1}$, this requirement reads
\begin{equation}
    \mathcal{T}\hat{L}_{j-1} = \hat{L}_j \, ,
\end{equation}
i.e.\ translating a jump operator by one site yields the next one,
so the spatial profile is the same around every center site.

To make this concrete, consider a Kitaev-like system in which the
annihilation operators take the Bogoliubov form
$\hat{a}_k = v_k \hat{c}_k + u_k \hat{c}^\dagger_{-k}$,
with parameters $u_k$ and $v_k$ to be specified by the model.
Transforming back to the real-space basis, the jump operators read
\begin{equation}
\label{eq:translation}
    \hat{L}_j
    = \sum_l \Bigl(\tilde{\mathbf V}_{jl}\,\hat{c}_l + \tilde{\mathbf U}_{jl}\,\hat{c}_l^\dagger\Bigr),
\end{equation}
where the coefficients are
\begin{equation}
\label{eq:prefactors}
    \tilde{\mathbf U}_{jl} = \frac{1}{\sqrt{N}}\sum_k e^{ikl} \mathbf B_{jk}\,u_k \, ,
    \qquad
    \tilde{\mathbf V}_{jl} = \frac{1}{\sqrt{N}}\sum_k e^{ikl} \mathbf B_{jk}\,v_k \, .
\end{equation}
The translation constraint $\mathcal{T}\hat{L}_{j-1} = \hat{L}_j$ requires
$\tilde{\mathbf U}_{jl} = \tilde{\mathbf U}_{j-1,\,l-1}$ and $\tilde{\mathbf V}_{jl} = \tilde{\mathbf V}_{j-1,\,l-1}$.
Substituting into~\cref{eq:prefactors}, this imposes
$e^{ik}\mathbf B_{jk} = \mathbf B_{j-1,k}$ for all $j, k$.
Denoting $\sqrt{N}\, \mathbf B_{0k} =: \beta_k$, the unique solution is
\begin{equation}
\label{eq:miximg:mat:pbc}
    \mathbf B_{jk} = \beta_k \mathrm e^{-\mathrm ikj}/\sqrt{N} \, ,
\end{equation}
which is precisely of the form of~\cref{eq:more:general:B} with the
unitary $\mathbf{U}_{jk} = \mathrm e^{-\mathrm ikj}/\sqrt{N}$.

This result has a significant implication: for translationally invariant
systems, the most general linear jump operators that respect the symmetry and prepare the ground state give rise to an analytically tractable spectrum. 
One can therefore make rigorous, general statements about the scaling of preparation times, which are fully determined by the single-mode weights $\beta_k$.
\section{Jump-Operator Localization in the Kitaev Chain}
\label{app:kitaev}
In the following we consider the Kitaev chain Hamiltonian~\cite{Kitaev2001}
\begin{equation}
\hat H=\sum_{j=1}^{N}\Big[
-t\,(\hat c_j^\dagger \hat c_{j+1}+\hat c_{j+1}^\dagger \hat c_j)
-\mu\Big(\hat c_j^\dagger \hat c_j-\frac12\Big)
+\Delta\,(\hat c_j \hat c_{j+1}+\hat c_{j+1}^\dagger \hat c_j^\dagger)
\Big].
\label{eq:kitaev_realspace}
\end{equation}
After a Fourier transformation it becomes 
\begin{equation}
    \hat H = \sum_k \epsilon_k \, \hat c_k^\dagger \hat c_k  + \Delta \Big( \mathrm e^{-\mathrm ik} \hat c_k \hat c_{-k} + \mathrm e^{\mathrm ik} \hat c_{-k}^\dagger \hat c_k^\dagger\Big) \, , 
\end{equation}
where $\epsilon_k = -\mu -2t\cos(k)$.
The Hamiltonian is already block diagonal in blocks $(-k,\, k)$, so we can directly consider a single block. The Hamiltonian can then be written in a simpler form by introducing the Nambu spinor $\hat \Psi_k = \big( \hat c_k, \, \hat c_{-k}^\dagger \big)^T$
\begin{equation}
    \hat H = \sum_{k>0} \hat \Psi_k^\dagger H_k \hat \Psi_k, \hspace{0.5cm} H_k = \begin{pmatrix}
        \epsilon_k & -i\Delta_k\\
        i\Delta_k & -\epsilon_k
    \end{pmatrix},
\end{equation}
where we introduced the new matrix $\hat H_k$ and the parameter $\Delta_k = 2\Delta \sin(k)$. Note that diagonalizing the Hamiltonian amounts to diagonalizing the matrix $\hat H_k$, since unitary transformations conserve the commutation relations of the Nambu spinor. 
To explore the connection to the topological properties of the system more, we can rewrite the Hamiltonian matrix in a more geometrical manner as
\begin{equation}
    \hat H_k = \mathbf d(k)\,  \hat{\bm \sigma}, \hspace{0.5cm} \mathbf d(k) = \big( 0, \, \Delta_k, \, \epsilon_k \big).
\end{equation}
Here we introduced the Pauli spin vector $\hat{\bm\sigma} = \big(\hat\sigma_x, \, \hat\sigma_y, \, \hat\sigma_z\big)^T$. 
The vector $\mathbf d(k)$ is a vector on the Bloch sphere and it prescribes an oval rotation in the $y$-$z$-plane around (0, 0, $-\mu$). Note that this rotation circles around the origin for $|\mu|<2t$, while at
$|\mu|=2t$ it passes through the origin.
For the sake of generality, we consider an arbitrary vector $\mathbf n(k) = \mathbf d(k)/|\mathbf d(k)| = \big( n_x, n_y, n_z \big)$. In this general case, the eigenvalues are given by $\pm |\mathbf d|$, and we can directly consider only the upper band in the following. Then, the eigenvector reads
\begin{equation}
    \mathbf v = e^{i\alpha(k)}\frac{1}{\sqrt{2(1-n_z)}} \begin{pmatrix}
        n_x - in_y \\
        1-n_z
    \end{pmatrix} .
    \label{eq:app:v:gauge}
\end{equation}
Note that we explicitly wrote the gauge freedom given by the phase
$\alpha(k)$. In the gauge of~\cref{eq:app:v:gauge}, the eigenvector $\mathbf{v}(k)$
develops a discontinuity in the topological phase $|\mu|<2t$ (for
$\Delta\neq0$). As we show below, this discontinuity can be
compensated by an appropriate gauge choice of $\alpha(k)= k/2$.

On the critical lines a simple compensation is not possible. To see this, we can make the singular structure at the Ising critical points
explicit. For definiteness, consider $\mu=2t$, where the gap closes at
$k=\pi$, and write $q=k-\pi$. For $\Delta\neq0$,
\begin{align}
    \epsilon_k &= -tq^2+O(q^4),\\
    \Delta_k &= -2\Delta q+O(q^3),\\
    |\mathbf d(k)| &= 2|\Delta||q|+O(|q|^3).
\end{align}
Consequently,
\begin{align}
    n_y(k)
    &= -\operatorname{sgn}(\Delta)\operatorname{sgn}(q)
       +O(q^2),\\
    n_z(k)
    &= -\frac{t}{2|\Delta|}|q|+O(|q|^3).
\end{align}
In the gauge of~\cref{eq:app:v:gauge}, the Bogoliubov coefficients therefore behave as
\begin{align}
    \mathbf v_1(k)
    &= \frac{i\,\operatorname{sgn}(\Delta)}{\sqrt{2}}
       \operatorname{sgn}(q)+O(|q|),\\
    \mathbf v_2(k)
    &= \frac{1}{\sqrt{2}}+O(|q|),
\end{align}
where $\mathbf v_{1/2}$ denote the entries in $\mathbf v$.
Thus one Bogoliubov coefficient has a sign discontinuity at the critical
momentum, while the other remains finite and continuous. It is therefore not possible to make both entries continuous by an appropriate choice of the gauge, and we chose $\alpha(k)=-k/4$.
The transition at $\mu=-2t$, where the gap closes at $k=0$, follows analogously.

\begin{figure}[t]
    \centering
    \begin{minipage}{0.55\textwidth}
        \centering
        \begin{overpic}[width=\textwidth]{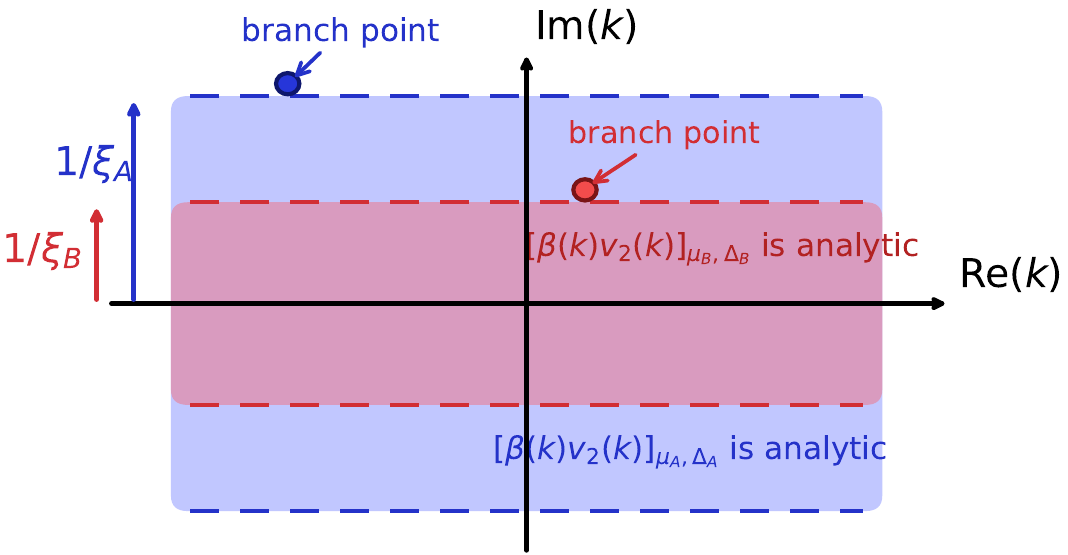}
            \put(1,55){\large\textbf{a)}}
        \end{overpic}
    \end{minipage}
    \hfill
    \begin{minipage}{0.44\textwidth}
        \centering
        \begin{overpic}[width=\textwidth]{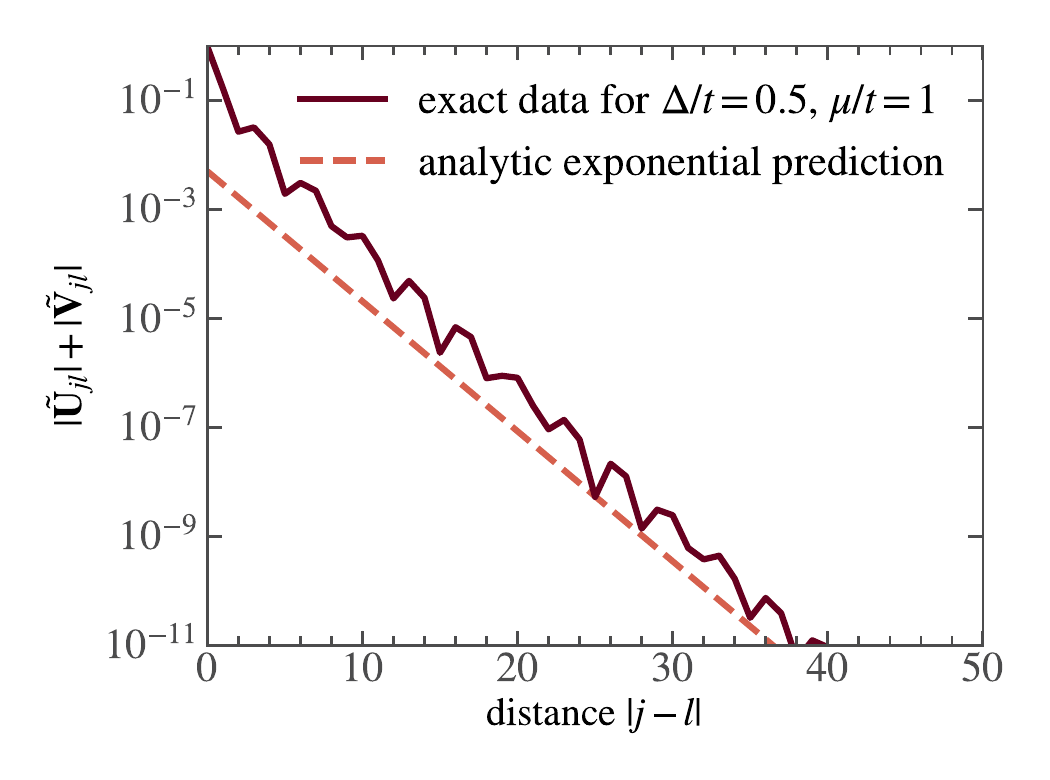}
            \put(8,72){\large\textbf{b)}}
        \end{overpic}
    \end{minipage}

    \caption{
    \textbf{Localization of the jump operators in the topological phase.}
    Left: Schematic analytic continuation to the complex momentum plane.
    For a given set of Hamiltonian parameters, the localization length $\xi$ is
    determined by the singularity closest to the real axis: the corresponding
    branch point bounds the strip of analyticity and lies at an imaginary distance
    $1/\xi$ from the real axis. 
    Right: Real-space jump-operator coefficients $|\tilde{\mathbf U}_{jl}| + |\tilde{\mathbf V}_{jl}|$,
    comparing the exact numerical Fourier coefficients with the asymptotic
    exponential decay rate derived in Eq.~\eqref{eq:app:cosh:xi} in the topological phase. The agreement confirms the
    exponential decay governed by $\xi$.
    }
    \label{fig:comp:exact:analytic:jump:ops}
\end{figure}

To diagonalize the full Hamiltonian, we can define the new annihilation operators $\hat \gamma_k = \mathbf v_1(k) \hat c_k + \mathbf v_2(k) \hat c_{-k}^\dagger$, bringing $\hat H$ into the form
\begin{equation}
    \hat H = \sum_k E_k \hat\gamma_k^\dagger \hat\gamma_k + \text{const.} \,, \hspace{0.5cm} E_k = |\mathbf d(k)|.
\end{equation}
Since $E(k)\geq 0$, the groundstate is the Bogoliubov vacuum $\hat\gamma_k \ket{\mathrm{vac}} = 0$ for every $k$. 
Knowing the diagonal annihilation operators we can rewrite the admissible jump operators as
\begin{align}
    \hat L_l = \sum_k \mathbf B_{lk} \hat \gamma_k = \sum_k \mathbf B_{lk} \frac{1}{\sqrt N}\sum_j \Big( \mathbf v_1(k) e^{ijk} \hat c_j + \mathbf v_2(k) e^{ijk} \hat c_j^\dagger\Big) = \sum_j \tilde{\mathbf V}_{lj} \hat c_j + \tilde{\mathbf U}_{lj} \hat c_j^\dagger,
\end{align}
where we can use \cref{eq:miximg:mat:pbc} to bring the prefactors into the form
\begin{equation}
    \tilde{\mathbf V}_{lj} = \frac{1}{N}\sum_k  \beta_k \mathbf v_1(k) \mathrm e^{\mathrm i(j-l)k} , \hspace{0.5cm} \tilde{\mathbf U}_{lj} = \frac{1}{N}\sum_k \beta_k \mathbf v_2(k) \mathrm e^{\mathrm i(j-l)k}.
\end{equation}
In the trivial phase $|\mu|>2t$ there exists an analytic continuation of the vector $\mathbf v$ with gauge $\alpha(k)=0$, meaning that a simple analytic $\beta_k$ will lead to an exponential behavior for $|l-j|\to \infty$.
In the topological phase, fixing the gauge $\alpha(k)=k/2$ leads again to a $\mathbf v$ featuring an analytic continuation to a strip in the complex plain, and a simple analytic $\beta_k$ once more leads to an exponential behavior for $|l-j|\to \infty$.
In the special case of the system parameters lying on the ellipse $\Delta^2+(\mu/2)^2=t^2$, note that $|\mathbf d| = 2t + \mu \cos k$ and by choosing
\begin{equation}
    \beta_k = \frac{|\mathbf d| \sqrt{2(1-n_z)}}{2\cos(k/2)} \; ,
    \label{eq:beta_on_the_ellipse}
\end{equation}
the prefactors take the form
\begin{equation}
    \beta_k \mathbf v = \mathrm e^{-\mathrm ik/2} \begin{pmatrix}
        -2\mathrm i \Delta \sin (k/2)\\
        (2t + \mu)\cos(k/2)\\
    \end{pmatrix}
    =  \begin{pmatrix}
        -\Delta (1-\mathrm e^{-\mathrm i k})\\
        \frac{2t+\mu}{2}(1+\mathrm e^{-\mathrm ik})
    \end{pmatrix}\; .
    \label{eq:app:beta:v:ellipse}
\end{equation}
Due to the particularly simple representation in terms of Fourier modes, the real space jump operators are given by
\begin{equation}
    \hat L_j = -\Delta\hat c_j  + \frac{2t+\mu}{2} \hat c_j^\dagger + \Delta \hat c_{j+1} + \frac{2t+\mu}{2} \hat c_{j+1}^\dagger \; .
    \label{eq:Lj:ellipse}
\end{equation}
We want to stress that especially in the case $\mu=0$, $\Delta=t$ the jump operators get particularly simple, and we recover the ones introduced in \cite{Diehl_2011} (up to the changed sign in $\Delta$ in the definition of $\hat H$).
We still need to show the analyticity of the Bogoliubov coefficients in the trivial and topological phase. Though tedious, this brings about a second advantage. 
For an arbitrary point in the topological phase, we need to make use of general theorems from complex analysis. 
First, we know that $1-n_z \geq 0$, and the zero (present in the topological phase) is of even order. Here, the chosen gauge is of particular importance, as it compensates the phase picked up by the eigenvector through the non-trivial winding. Then, the periodic continuation of $\mathbf v$ has an analytic continuation to a strip around the real axis.
By Theorem 5 in~\cite{boyd2001chebyshev}, the matrices $\tilde{\mathbf V}_{lj}$ and $\tilde{\mathbf U}_{lj}$ decay exponentially $\sim e^{-|l-j|/\xi}$. This behavior is visible from the Fourier coefficients in \cref{fig:comp:exact:analytic:jump:ops} b).

Moreover, it is actually possible to calculate the exact jump operator localization length $\xi$, as it is given by 
\begin{equation}
    1/\xi = \sup\{\, v\,  | \, \beta_{(\cdot)} \mathbf v_2(\cdot): C_v \to \mathds C \; \text{is} \; \text{analytic}\} \, ,
    \label{eq:app:localization:lenght}
\end{equation}
where we defined the complex strip $C_v = \{\mathds R + \mathrm ia| \, |a|<v\}$. 
The reason for the coefficients not being analytic on the whole of $\mathds C$ is the branch cut of the complex square root at the negative reals. This plays a role in the definition of $n_z$ through $|\mathbf d|$.
To avoid confusion, in the following we will denote $f=|\mathbf d|^2$.
Since $f>0$ in the reals, the branch cut happens on the lines where $\mathrm{Im} f = 0$.
Starting from the reals and going along the zero imaginary part lines, we encounter the branch cut exactly at $\mathrm{Re} f=0$, if the zero crossing is of odd multiplicity. 
The width of analyticity is thus given by the first (odd) zero of $f$: $f=0$. 
This connection between momentum-space analyticity and real-space
localization is illustrated schematically in~\cref{fig:comp:exact:analytic:jump:ops}a).
The singularity closest to the real axis determines the boundary of the
analyticity strip and hence the inverse localization length $1/\xi$.
Different positions of the nearest branch point therefore directly correspond
to different localization lengths.
We can analytically extract these zeros by calculating the real and imaginary parts of $f$. After some straightforward algebra we get
\begin{align}
    \text{Re} f(u + \mathrm iv) &= \mu^2 + 2(\Delta^2 + t^2) + 2(t^2 - \Delta^2) \cos(2u)\cosh(2v) + 4t\mu \cos(u) \cosh(v)\, ,\\
    \text{Im}f(u + \mathrm i v) &= -2(t^2 - \Delta^2) \sin(2u)\sinh(2v) - 4t\mu \sin(u)\sinh(v)\, .
\end{align}
First, we will determine the lines on which $\mathrm{Im} f=0$ is satisfied.
There are two trivial solutions, (I) $v=0$ and (II) $u=\pi n$. The first one is not of interest, while the second one could give rise to a zero in the complex plain. Dividing by $\sin(u)\sinh(v)$ and rearranging gives the third (III) solution 
\begin{equation}
    \cos(u)\cosh(v) = \frac{t\mu}{2(\Delta^2-t^2)} \, .
\end{equation}
Note that this also entails that possible values for $v$ on this curve lie in the interval $[v_\text{min}^\text{(III)}, \infty)$ with $v_{\min}^{(\mathrm{III})}
=
\operatorname{acosh}\!\left[
\max\!\left(
1,\frac{t\mu}{2(t^2-\Delta^2)}
\right)
\right]$ (which we will see is necessary to have zero crossings in this case).
We first analyze (II). For definiteness, we take $\mu>0$; for $\mu<0$ the same analysis applies with even and odd $n$ interchanged. Inserting into the real part and setting it to zero gives
\begin{equation}
0=\mu^2+4\Delta^2+(-1)^n4t\mu\cosh(v)
+4(t^2-\Delta^2)\cosh^2(v).
\end{equation}
Since the sign of the prefactor in the second term is completely determined by $n$, the zero crossing for even $n$ (if existing) lies at higher $v$ than the one for odd $n$. Thus, we can directly restrict to odd $n$ and find the zeros
\begin{equation}
v_{\pm}=\operatorname{acosh}(x_{\pm}),\qquad
x_{\pm}=
\frac{t\mu\pm2\Delta\sqrt{(\mu/2)^2+\Delta^2-t^2}}
{2(t^2-\Delta^2)}.
\end{equation}
Note that the zero crossing exists exactly if the parameters are outside of the ellipse. Exactly on the ellipse, the zero crossing is of second order, and the function is positive. Furthermore, $x_-\geq1$, so $v_-$ is a genuine zero crossing and gives the smallest zero of solution (II).
To see this, define
\begin{equation}
Q=t^2-\Delta^2,
\qquad
S=(\mu/2)^2+\Delta^2-t^2>0.
\end{equation}
For $\Delta\neq t$, we can write
\begin{equation}
x_--1
=
\frac{t\mu-2Q-2\Delta\sqrt{S}}{2Q}.
\end{equation}
Using
\begin{equation}
\left(t\mu-2Q\right)^2-4\Delta^2S
=
Q(\mu-2t)^2,
\end{equation}
we obtain
\begin{equation}
x_--1
=
\frac{(\mu-2t)^2}
{2\left(t\mu-2Q+2\Delta\sqrt{S}\right)}
\geq0.
\end{equation}
The case $\Delta=t$ follows by continuity. Hence $x_-\geq1$, with equality only at the critical point $\mu=2t$.

We still need to analyze solution (III). Inserting into the real part and setting to zero, we get
\begin{equation}
    0 = \frac{t^2\mu^2}{\Delta^2 - t^2} + (\mu^2 + 4t^2) \cosh(v)^2 - 4(t^2 - \Delta^2) \cosh(v)^4 \, .
\end{equation}
Solving for the zeros we get
\begin{equation}
    v_\pm = \mathrm{acosh}(\sqrt{x_\pm})\, , \quad x_\pm = \frac{-(\mu^2 + 4t^2) \pm (\mu^2 - 4t^2)}{8(\Delta^2-t^2)} \; .
\end{equation}
We can further analyze these solutions. Both solutions only exist if $t>\Delta$. Additionally, they are only visible if they satisfy $v_\pm \geq v_\text{min}^\text{(III)}$. We can analyze this conditions in the two cases. 
For $v_-$ this condition would require
$\sqrt{t^2-\Delta^2}\geq t$, which is impossible for $\Delta\neq0$. 
For $v_+$ on the other hand, the condition simplifies to $t^2\geq (\mu/2)^2 + \Delta^2$, meaning that the parameters must lie inside of the ellipse. 
If they are exactly on the ellipse, the zero crossing is exactly at the edge of the curve and there is no branch cut. 
In total, we thus find the following analytical behavior
\begin{equation}
    \cosh(\xi^{-1}) = \begin{cases}
        \sqrt{\frac{t^2}{t^2-\Delta^2}}\, , \mathrm{inside \; the \; ellipse},\\
        \frac{t\mu - 2\Delta \sqrt{(\mu/2)^2 +\Delta^2 - t^2}}{2(t^2 - \Delta^2)} \, , \mathrm{outside\; the \; ellipse}.
    \end{cases}
    \label{eq:app:cosh:xi}
\end{equation}
Now we are able to investigate the behavior close to the critical points. First let's consider $\Delta \to 0$ (XX-criticality). In this case we have in leading order
\begin{equation}
    \xi^{-1} = \text{acosh}\Big(1 + \frac{1}{2} (\Delta/t)^2\Big) = \Delta/t + \mathcal O\big(  (\Delta/t)^2\big) \; .
\end{equation}

We can similarly approach the Ising critical points at $\mu=\pm 2t$
for fixed $\Delta\neq 0$. For definiteness, consider $\mu\to 2t^{-}$
and define $\delta\mu=2t-\mu$. Close to the transition we are outside
the ellipse, and expanding~\cref{eq:app:cosh:xi} gives
\begin{equation}
    \cosh(\xi^{-1})
    =
    1+\frac{(\delta\mu)^2}{8\Delta^2}
    +\mathcal{O}(\delta\mu^3).
\end{equation}
Using $\operatorname{acosh}(1+x)\simeq \sqrt{2x}$ for $x\ll 1$, we
therefore obtain
\begin{equation}
    \xi^{-1}
    =
    \frac{\delta\mu}{2|\Delta|}
    +\mathcal{O}(\delta\mu^2),
\end{equation}
and hence
\begin{equation}
    \xi
    \sim
    \frac{2|\Delta|}{|2t-\mu|}.
\end{equation}
Thus, the jump-operator localization length diverges upon approaching
the Ising critical point, with the Ising critical exponent $\nu=1$.
The transition at $\mu=-2t$ follows analogously, with the corresponding
gap closing occurring at $k=0$.

A qualitatively different behavior is found when approaching the
ellipse from either side. Define the value of $\Delta$ on the ellipse as
\begin{equation}
    \Delta_{\rm e}
    =
    \sqrt{t^2-(\mu/2)^2}.
\end{equation}
Taking the limit $\Delta\to\Delta_{\rm e}$ in either branch of ~\cref{eq:app:cosh:xi} gives
\begin{equation}
    \lim_{\Delta\to\Delta_{\rm e}^{\pm}}
    \cosh(\xi^{-1})
    =
    \frac{2t}{|\mu|},
\end{equation}
and therefore, for $0<|\mu|<2t$,
\begin{equation}
    \lim_{\Delta\to\Delta_{\rm e}^{\pm}}
    \xi
    =
    \frac{1}{
    \operatorname{acosh}\!\left(2t/|\mu|\right)
    }.
    \label{eq:xi_ellipse_limit}
\end{equation}
Importantly, this finite limiting value is not the localization length
exactly on the ellipse, when choosing \cref{eq:beta_on_the_ellipse}. As shown in~\cref{eq:app:beta:v:ellipse} and~\cref{eq:Lj:ellipse}, on the ellipse
the momentum-space jump operators reduce to finite trigonometric
polynomials and the corresponding real-space jump operators are
strictly finite range. Equivalently, in terms of~\cref{eq:app:localization:lenght}, the
analyticity strip has unbounded width, so that
$\xi^{-1}=\infty$, or $\xi=0$.
Consequently, for generic points on the ellipse with
$0<|\mu|<2t$, the localization length is discontinuous: approaching
the ellipse from either side yields the finite value in
Eq.~\eqref{eq:xi_ellipse_limit}, whereas directly on the ellipse the
jump operators become strictly local and $\xi=0$. The point
$\mu=0$, $\Delta=t$ is a special case, for which the limiting value
itself approaches zero.
Note that also the additional algebraic factor in the coefficient decay can be extracted using Cauchy's formula.

\section{Mpemba-type accelerated state preparation}
\label{app:mpemba}

As discussed in the main text and in App.~\ref{app:sec:spectral},
the slowest Lindbladian eigenmodes correspond to coherences between
states of opposite fermion parity and are therefore odd under
fermion-parity conjugation. Such coherences are excluded from physical
fermionic density matrices by fermion-parity superselection~\cite{Wick1952aa,PhysRevA.104.032411}. Consequently,
physical fermionic states have vanishing overlap with these modes, and
their asymptotic relaxation is governed by the next decay rate,
$2\Delta_{\mathcal L}$.
This can be understood directly from the spectral decomposition
\begin{equation}
    \hat\rho(t)-\hat\rho_{\rm ss}
    =
    \sum_{\alpha} c_{\alpha}
    e^{\lambda_{\alpha} t}
    \hat R_{\alpha},
    \label{eq:mpemba:spectral}
\end{equation}
where $\hat R_{\alpha}$ are right Lindbladian eigenmodes and the
coefficients $c_{\alpha}$ are fixed by the initial state.
The vanishing of the coefficient associated with the slowest
Lindbladian mode is precisely the spectral mechanism underlying a
strong Mpemba effect~\cite{Lu2017,Kumar2020,TEZA20261}: the asymptotic relaxation is then
controlled by a faster mode.

\begin{figure}[t]
    \centering
    \includegraphics[width=0.65\columnwidth]{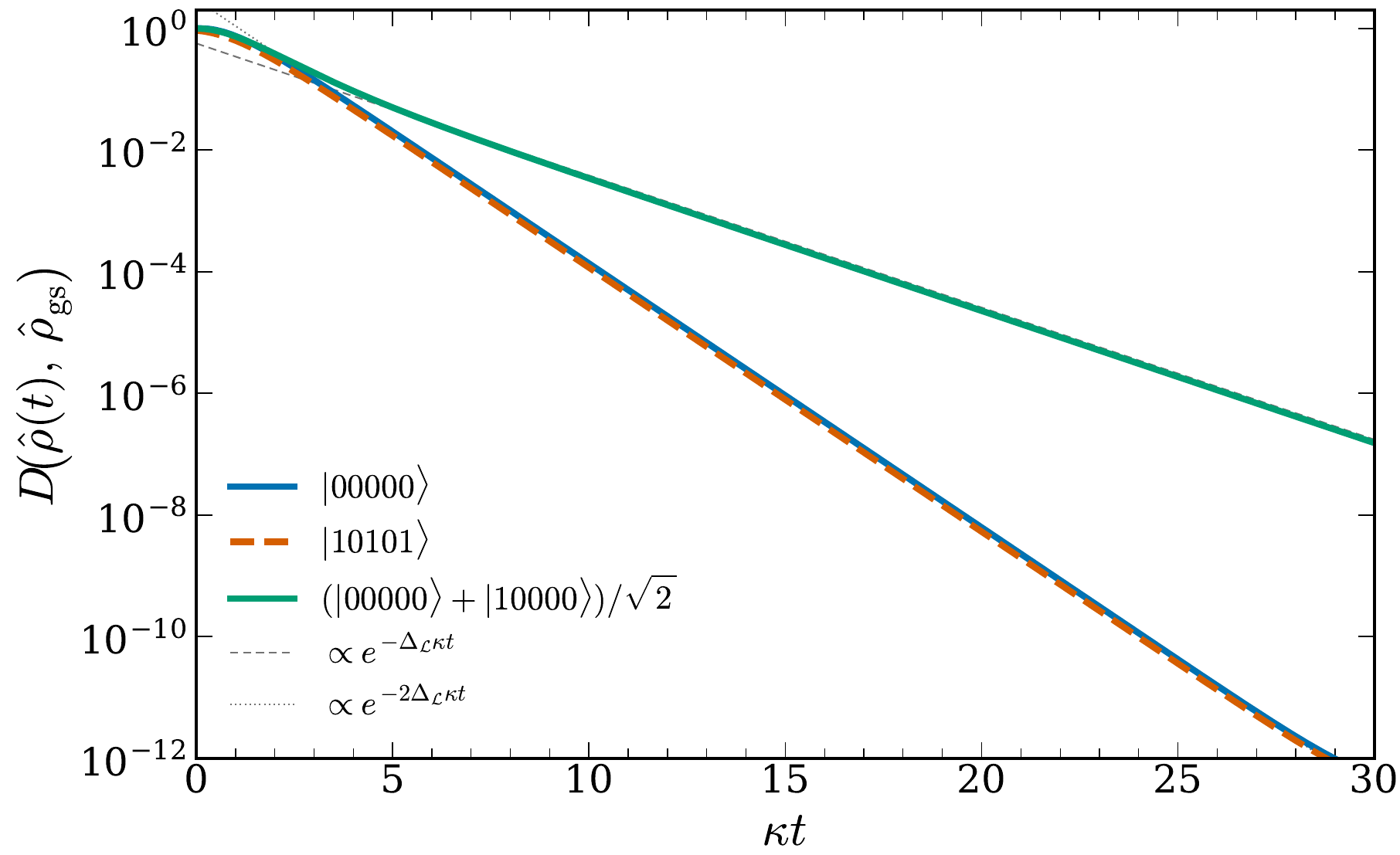}
    \caption{
    \textbf{Parity-controlled relaxation and Mpemba-type acceleration.}
    Trace distance $D(\hat\rho(t),\hat\rho_{\rm gs})$ from the target
    steady state for a Kitaev chain with $N=5$, $t=1$, $\Delta=0$, and
    $\mu=1$. The full Lindblad dynamics,
    including the Kitaev Hamiltonian and Bogoliubov jump operators
    $\hat L_k=\sqrt{\kappa}\,\hat a_k$, is solved by exact diagonalization
    of the Lindbladian. The definite-parity states
    $|00000\rangle$ (blue) and $|10101\rangle$ (orange dashed) have no
    overlap with the slowest odd-parity coherence modes and asymptotically
    decay with rate $2\Delta_{\mathcal L}$. The parity-coherent state
    $(|00000\rangle+|10000\rangle)/\sqrt{2}$ (green) accesses these modes
    and instead decays with rate $\Delta_{\mathcal L}$. Grey reference
    lines indicate the corresponding asymptotic decay rates. Time is shown
    in units of $\kappa^{-1}$.
    }
    \label{fig:mpemba_parity}
\end{figure}

\Cref{fig:mpemba_parity} illustrates this mechanism explicitly at a representative point on the XX critical line (for the phase diagram, see Fig. 2 in the main text).
The coherent superposition of states with opposite fermion parity contains
coherences between the corresponding parity sectors and therefore accesses
the slowest odd-parity Lindbladian modes, relaxing asymptotically at the rate
$\Delta_{\mathcal L}$. By contrast, the two definite-parity initial states
have vanishing overlap with these modes and consequently relax faster, at the
rate $2\Delta_{\mathcal L}$.
While these parity coherences are forbidden for physical fermionic
density matrices, they are allowed in spin or hard-core-boson
realizations, where no analogous fermion-parity superselection rule
applies. In such systems, the spectral structure therefore identifies
initial states with vanishing overlap with the slowest modes and hence
provides a strong-Mpemba-type factor-of-two acceleration of dissipative
state preparation~\cite{Westhoff2025,Beato26}.
Larger system sizes can be accessed using covariance-matrix methods for quadratic dynamics, including the Lyapunov-equation approach reviewed in~\cite{Lacerda2026thesis}.

\section{Nonmonotonic behavior in the fidelity}
\begin{figure}[t]
    \centering
    \begin{minipage}{0.48\textwidth}
        \centering
        \begin{overpic}[width=\textwidth]{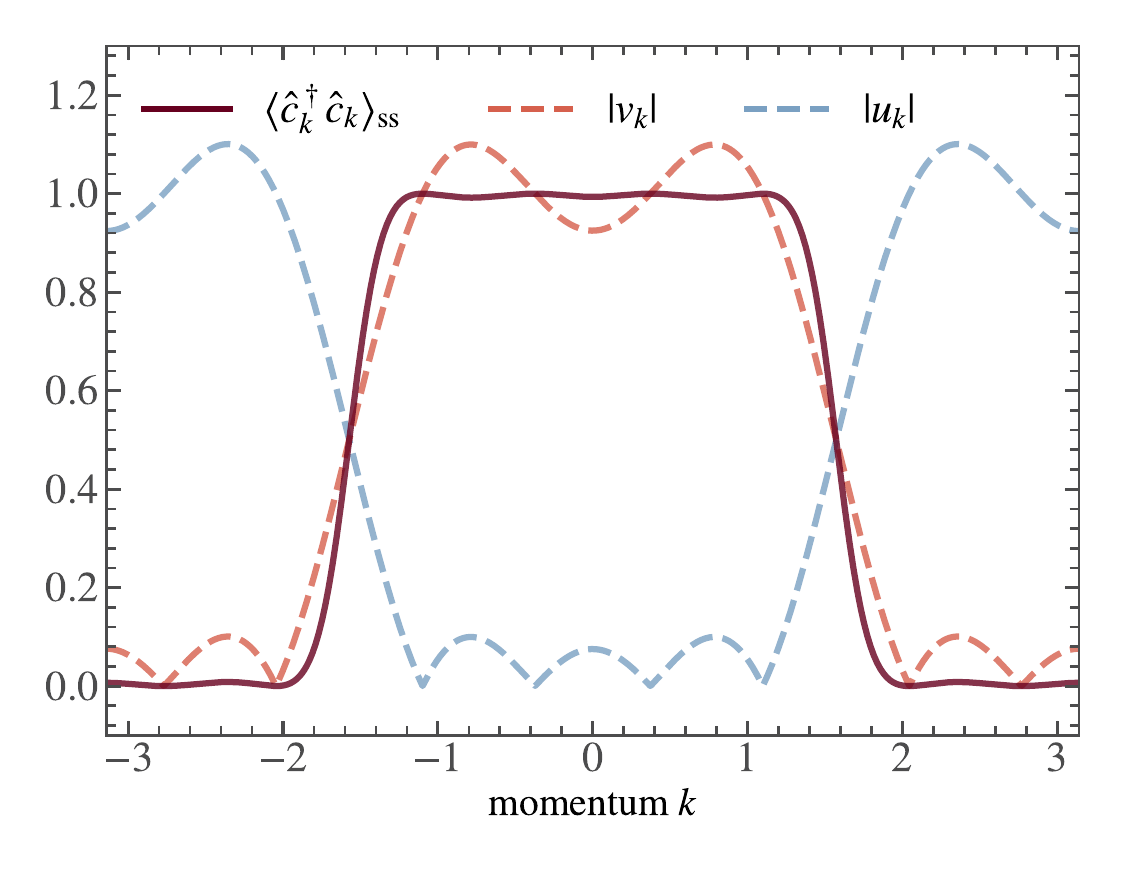}
            \put(1,75){\large\textbf{a)}}
        \end{overpic}
    \end{minipage}
    \hfill
    \begin{minipage}{0.48\textwidth}
        \centering
        \begin{overpic}[width=\textwidth]{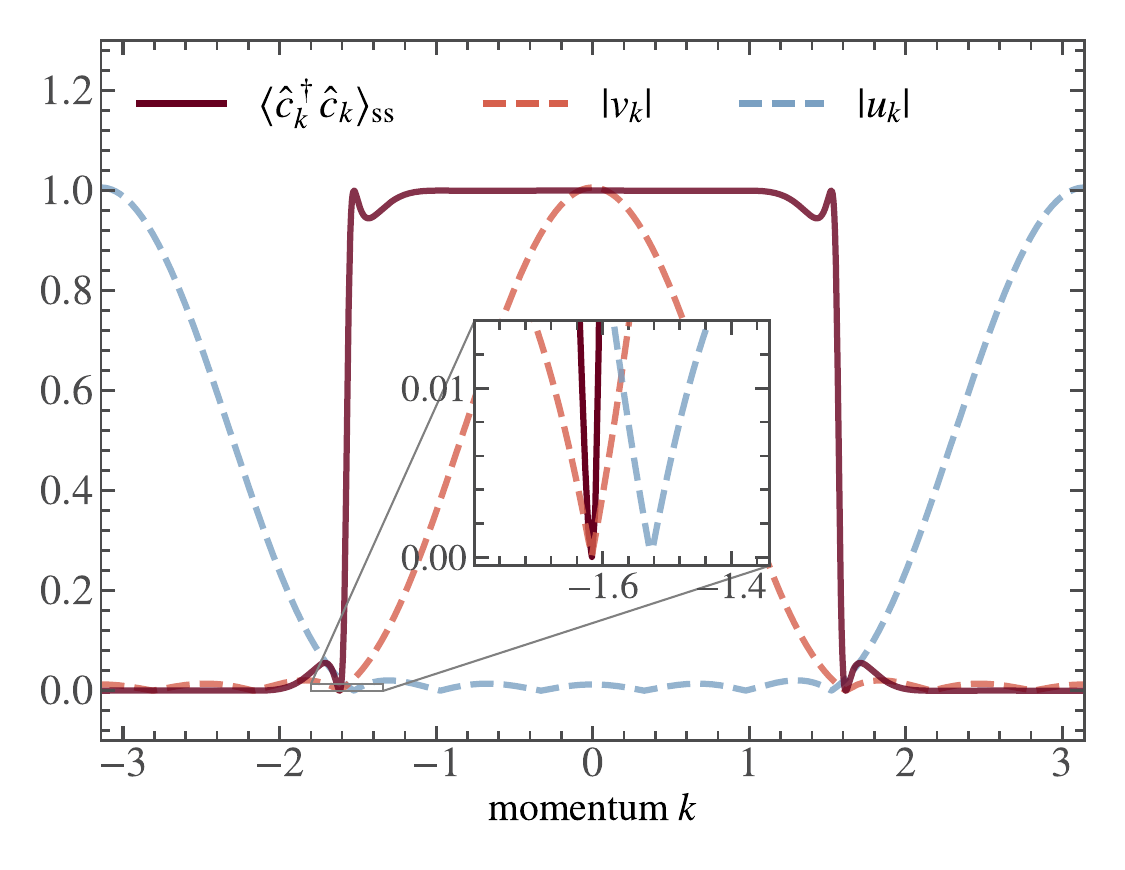}
            \put(0,75){\large\textbf{b)}}
        \end{overpic}
    \end{minipage}

    \caption{
    \textbf{Momentum population at XX-criticality.}
    We show the momentum population \cref{eq:momentum:population} (red) and the corresponding jump operator prefactors $v_k$ (orange) and $u_k$ (blue) from \cref{eq:truncated:prefactors} for $\beta_k = |\cos(k)|^n$ with exponents $n=0$ (subfigure a) and $n=1.65$ (subfigure b).
    }
    \label{fig:k:population:small:n}
\end{figure}

\begin{figure}[t]
    \centering
    \begin{minipage}{0.48\textwidth}
        \centering
        \begin{overpic}[width=\textwidth]{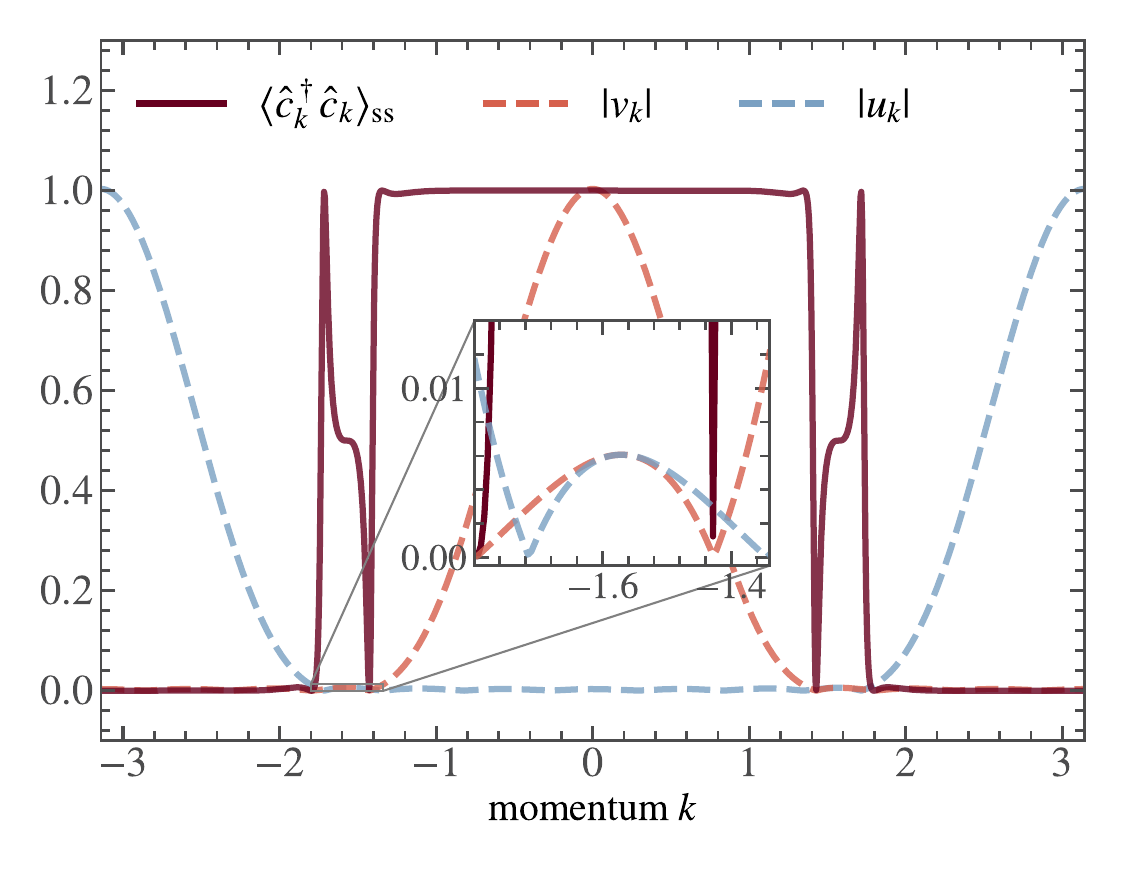}
            \put(1,75){\large\textbf{a)}}
        \end{overpic}
    \end{minipage}
    \hfill
    \begin{minipage}{0.48\textwidth}
        \centering
        \begin{overpic}[width=\textwidth]{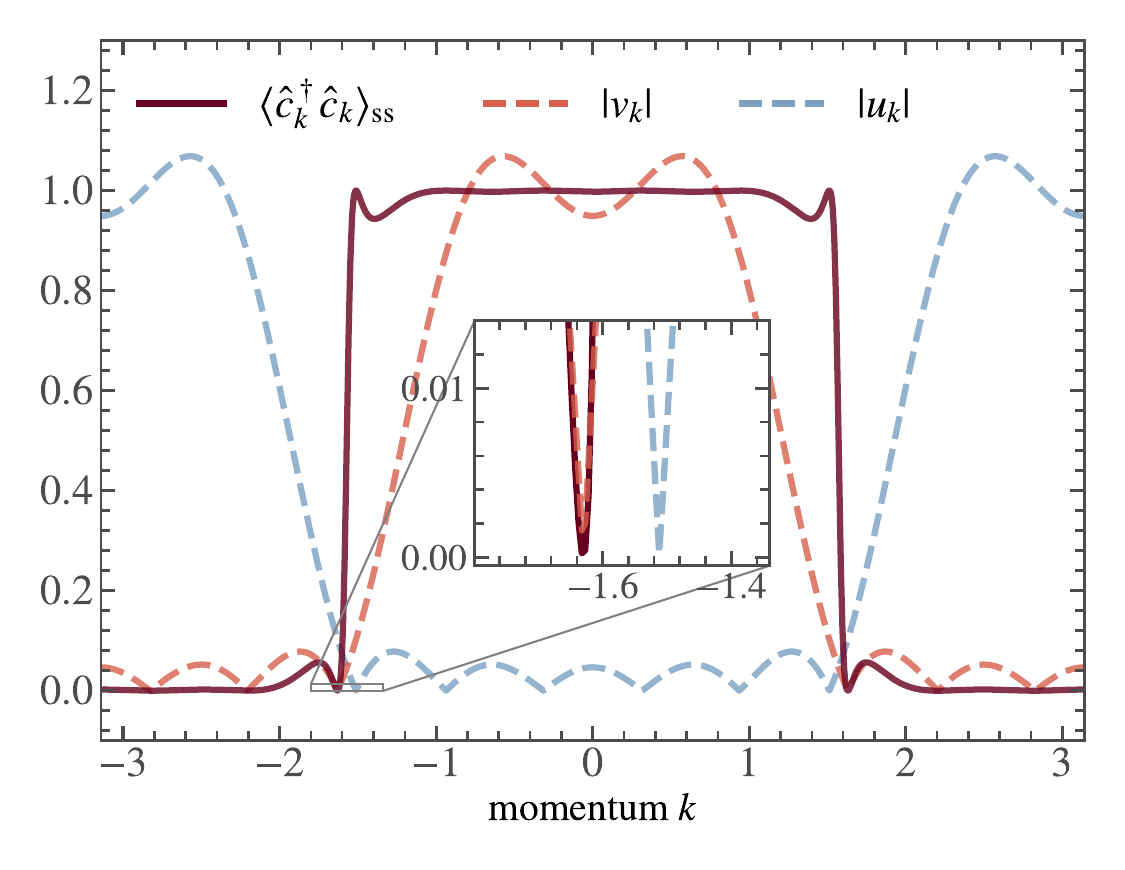}
            \put(0,75){\large\textbf{b)}}
        \end{overpic}
    \end{minipage}

    \caption{
    \textbf{Momentum population at XX-criticality for high tradeoff exponent $n=3$.}
    We show the momentum population \cref{eq:momentum:population} (red) and the corresponding jump operator prefactors $v_k$ (orange) and $u_k$ (blue) from \cref{eq:truncated:prefactors} for $\beta_k = |\cos(k)|^3$ (subfigure a) and $\beta_k'= 1-(1-\beta_k)^{16}$ (subfigure b).
    }
    \label{fig:k:population:large:n}
\end{figure}

The inset of Fig. 3 in the main text shows that the infidelity of the steady state first decreases upon increasing the tradeoff exponent $n$ and then increases again for $n \gtrsim 1.7$. Here we explain the origin of this non-monotonic behavior.
We can explicitly calculate the particle number in a specific momentum mode. For generality we consider Bogoliubov quasiparticles as jump operators $ \hat{L}_k = \hat\gamma_k = u_k \hat c_k + v_k\hat c_{-k}^\dagger$. Then, the steady-state populations are given by 
\begin{equation}
    \langle \hat c_k^\dagger \hat c_k\rangle_\mathrm{ss} = \frac{|v_k|^2}{|u_k|^2+|v_k|^2}\, .
    \label{eq:momentum:population}
\end{equation}
Note that this holds also for truncated jump operators, where $u_k$ and $v_k$ are not the Bogoliubov prefactors \cref{eq:app:v:gauge} but rather
\begin{equation}
    u_k = \sum_g \beta_g \mathbf v_1(g)f_R(k-g)\, , \quad v_k = \sum_g \beta_g \mathbf v_2(g)f_R(k-g) \, ,
    \label{eq:truncated:prefactors}
\end{equation}
i.e. the folding of the Bogoliubov prefactors with the function 
\begin{equation}
    f_R(g) = \mathrm e^{iRg} \frac{1-\mathrm e^{-i(2R+1)g}}{1-\mathrm e^{-ig}} \, .
\end{equation}
Here, $R$ denotes the spatial range each jump operator should have after the truncation of its tails, i.e. jump operator $\hat L_l$ is supported on $[l-R, l+R]$. 
Notice also that the populations in the momentum basis completely determine the particle-hole correlator in the real space basis
\begin{equation}
    \langle \hat c_j^\dagger \hat c_{j+r} \rangle_\mathrm{ss} = \frac{1}{N} \sum_k \mathrm e^{-irk}\langle \hat c_k^\dagger \hat c_k\rangle_\mathrm{ss} \, .
\end{equation}
Since this is a Fourier series, the regularity of the momentum space populations dictates the decay of the correlator with $r$. 
Let's now consider a specific point on the XX critical line, where $\mu=0$.
Then the momentum space populations also control the fidelity of the steady state compared to the target groundstate. 
We can calculate this fidelity in each $k$-sector alone, but let's first consider the $|k|<\pi/2$. The groundstate in each of these sectors is given by $\ket{\mathrm{GS}(k)} = \hat c_k^\dagger \hat c_{-k}^\dagger \ket{0}$, while the steady state is the pure state of $\ket{\mathrm{SS}(k)} = \frac{1}{\mathcal N}(u_k+v_k\hat c_k^\dagger \hat c_{-k}^\dagger)\ket{0}$, and the fidelity reads
\begin{equation}
    |\braket{\mathrm{GS}(k)|\mathrm{SS}(k)}| = \frac{|v_k|}{\sqrt{|u_k|^2 + |v_k|^2}} = \sqrt{\langle \hat c_k^\dagger \hat c_k\rangle_\mathrm{ss}},
\end{equation}
and similarly for $|k|>\pi/2$
\begin{equation}
    |\braket{\mathrm{GS}(k)|\mathrm{SS}(k)}| = \frac{|u_k|}{\sqrt{|u_k|^2 + |v_k|^2}} = \sqrt{1-\langle \hat c_k^\dagger \hat c_k\rangle_\mathrm{ss}}\,.
\end{equation}
Due to the tensor product structure of both the steady state as well as the groundstate, the full fidelity is calculated as 
\begin{equation}
    |\braket{\mathrm{GS}|\mathrm{SS}}| = \prod_{k>0} |\braket{\mathrm{GS}(k)|\mathrm{SS}(k)}|\,.
\end{equation}
This reveals that large deviations in the momentum space populations from 1 (for $|k|<\pi/2$) and 0 (for $|k|>\pi/2$) lead to large deviations in the fidelity.
When comparing truncated with non-truncated jump operators, whether the decay is accurately captured by the truncated ones now heavily depends on the function $\beta_k$ as it contributes non-trivially in \cref{eq:truncated:prefactors}.
To illustrate this behavior, we consider the XX-criticality ($\mu=0$) with $\beta_k = |\cos(k)|^n$, where we fix the truncation cutoff $R=5$ and vary the tradeoff exponent $n$ (same as in Fig. 3). 
We first consider the two exponents $n=0$ and $n=1.65$ in \cref{fig:k:population:small:n}. 
For $n=0$ (\cref{fig:k:population:small:n} (a)) we see that both $v_k$ and $u_k$ deviate strongly from the non-truncated curve (not shown, given by two boxes), leading to smoothened out edges at $k=\pm \pi/2$. 
In contrast, for $n=1.65$ (\cref{fig:k:population:small:n} (b)) the points at which $v_k$ and $u_k$ hit zero  moved much closer to $k=\pm \pi/2$ than for $n=0$. Also, the deviations in the regime where the parameters should be zero in the non-truncated case ($k<-\pi/2$ and $k>+\pi/2$) are much smaller than for the $n=0$ case. 
This also leads to much sharper edges at $k=\pm\pi/2$.
In contrast, for $n=3$ in \cref{fig:k:population:large:n} (a) the strong suppression due to the zero of third order around $k=\pm \pi/2$ leads to both $u_k$ and $v_k$ being very small around these points. Since the population depends only on their ratio, the truncation errors introduced by the cutoff are amplified strongly in the population, leading to a residual contribution around the jumps. 
Note however that this problem originates from the low value of $\beta_k\sim a|k|^n$ around the discontinuity at $k=\pm\pi/2$ of the Bogoliubov coefficients, and not from the high tradeoff exponent $n$ itself ($a=1$ for the particular choice of $\beta_k$). We can alleviate this problem, for instance, by strongly increasing the prefactor $a$ in the expansion of $\beta_k$ around $k=\pm \pi$. 
This is precisely what is shown in \cref{fig:k:population:large:n} (b), where we change the prefactor from $a=1$ to $a=16$ by considering $\beta'_k = 1-(1-\beta_k)^{16}$. This changes the behavior drastically and the population is much closer to the box shape corresponding to the non-truncated case.

\FloatBarrier
\bibliography{Literature}